\documentclass[fleqn,usenatbib]{mnras}

\usepackage{newtxtext,newtxmath}

\usepackage[T1]{fontenc}

\DeclareRobustCommand{\VAN}[3]{#2}
\let\VANthebibliography\thebibliography
\def\thebibliography{\DeclareRobustCommand{\VAN}[3]{##3}\VANthebibliography}

\usepackage{graphicx}	
\usepackage{amsmath}	
\usepackage{pifont}
\usepackage[dvipsnames]{xcolor}
\usepackage{chemformula}

\defcitealias{BR2006}{BR06}
\defcitealias{Kr2009}{KMT09}

\title[Molecular gas and pristine feedback]{Impact of H$_{\rm 2}$-driven star formation and stellar feedback from low-enrichment environments on the formation of spiral galaxies}

\author[M. Valentini et al.]
{Milena Valentini$^{1,2,3}$\thanks{Alexander von Humboldt Research Fellow\newline E-mail: valentini@usm.lmu.de}, 
Klaus Dolag$^{1,2,4}$, 
Stefano Borgani$^{5,3,6,7}$, 
Giuseppe Murante$^{3,7}$, 
Umberto Maio$^{3}$, 
\newauthor
Luca Tornatore$^{3}$, 
Gian Luigi Granato$^{3,8,7}$, 
Cinthia Ragone-Figueroa$^{8,3}$, 
Andreas Burkert$^{1,2,9}$, 
\newauthor
Antonio Ragagnin$^{10,3,7}$ and
Elena Rasia$^{3}$
\\ ~ \\
$^{1}$ Universit{\"a}ts-Sternwarte, Fakult{\"a}t f{\"u}r Physik,  Ludwig-Maximilians Universit{\"a}t  M{\"u}nchen, Scheinerstr. 1, D-81679 M{\"u}nchen, Germany\\
$^{2}$ Excellence Cluster ORIGINS, Boltzmannstr. 2, D-85748 Garching, Germany\\
$^{3}$ INAF - Osservatorio Astronomico di Trieste, via Tiepolo 11, I-34131 Trieste, Italy\\
$^{4}$ Max Planck Institute for Astrophysics, Karl-Schwarzschild-Str. 1, D-85741 Garching, Germany\\
$^{5}$ Astronomy Unit, Department of Physics, University of Trieste, Via Tiepolo 11, I-34131 Trieste, Italy\\
$^{6}$ INFN - National Institute for Nuclear Physics, Via Valerio 2, I-34127 Trieste, Italy\\
$^{7}$ IFPU - Institute for Fundamental Physics of the Universe, Via Beirut 2, I-34014 Trieste, Italy\\
$^{8}$ Instituto de Astronom\'ia Te\'orica y Experimental (IATE), Consejo Nacional de Investigaciones Cient\'ificas y T\'ecnicas de la\\ Rep\'ublica Argentina (CONICET), Universidad Nacional de C\'ordoba, Laprida 854, X5000BGR C\'ordoba, Argentina\\
$^{9}$ Max-Planck Institute for Extraterrestrial Physics, Giessenbacherstr. 1, 85748 Garching, Germany\\
$^{10}$ Dipartimento di Fisica e Astronomia "Augusto Righi", Alma Mater Studiorum Universit\`a di Bologna, via Gobetti 93/2, I-40129 Bologna, Italy\\
}

\date{Accepted 2022 July 25. Received 2022 July 22; in original form 2022 May 13}

\pubyear{2022}

\begin{document}
\label{firstpage}
\pagerange{\pageref{firstpage}--\pageref{lastpage}}
\maketitle

\begin{abstract}
The reservoir of molecular gas (H$_{\rm 2}$) represents the fuel for the star formation (SF) of a galaxy. 
Connecting the star formation rate (SFR) to the available H$_{\rm 2}$ is key to accurately model SF in cosmological simulations of galaxy formation.  
We investigate how modifying the underlying modelling of H$_{\rm 2}$ and the description of stellar feedback in low-metallicity environments (LMF, i.e. low-metallicity stellar feedback) in cosmological, zoomed-in simulations of a Milky Way-size halo influences the formation history of the forming, spiral galaxy and its final properties. 
We exploit two different models to compute the molecular fraction of cold gas (f$_{\rm H_{\rm 2}}$): $i)$ the theoretical model by Krumholz et al. (2009b) and $ii)$ the phenomenological prescription by Blitz \& Rosolowsky (2006). 
We find that the model adopted to estimate f$_{\rm H_{\rm 2}}$ plays a key role in determining final properties and in shaping the morphology of the galaxy. The clumpier interstellar medium (ISM) and the more complex H$_{\rm 2}$ distribution that the Krumholz et al. (2009b) model predicts result in better agreement with observations of nearby disc galaxies. This shows how crucial it is to link the SFR to the physical properties of the star-forming, molecular ISM.
The additional source of energy that LMF supplies in a metal-poor ISM is key in controlling SF at high redshift and in regulating the reservoir of SF across cosmic time. Not only is LMF able to regulate cooling properties of the ISM, but it also reduces the stellar mass of the galaxy bulge.
These findings can foster the improvement of the numerical modelling of SF in cosmological simulations.  
\end{abstract}





\section{Introduction} 
\label{sec:introduction}

Star formation (SF) is an essential process for the formation and evolution of cosmic structures.  Gas accreted from the large-scale structure provides the reservoir for SF, as soon as it cools radiatively while reaching the innermost regions of forming structures, and it meets temperature and density conditions to trigger SF. At that point, molecular hydrogen (H$_{2}$) in the densest and coldest phase of the interstellar medium (ISM) is converted into stars within dusty, giant molecular clouds (GMCs) \citep[see][for reviews]{Kennicutt1998araa, McKeeOstriker2007}. 

The cold ISM \citep[e.g.][for a recent review]{Saintonge2022} is the gas phase which plays the major role in determining the star formation rate (SFR) of a galaxy. Indeed, this quantity is regulated by the available mass of cold gas (along with gas infall and ejection,  e.g. galactic fountains), by the fraction of that gas which is in the molecular phase, and by the efficiency or rate at which H$_{2}$ is depleted and converted into stars. The aforementioned factors are highly dependent on redshift and on galaxy properties, e.g. the galaxy offset from the main sequence \citep[][and references therein]{Tacconi2020}.

Observational evidence suggests that the SFR correlates with the total (i.e. neutral hydrogen + H$_{2}$) gas surface density, albeit the correlation is even tighter when the H$_{2}$ alone is considered \citep[][]{WongBlitz2002,  Kennicutt2007,  Bigiel2008,  Leroy2008}. These works also show that the SFR correlates little to nothing with the neutral phase at low surface densities (although \citet{Bacchini2019} claim that the total gas volume density likely correlates better with the galaxy SFR in low-density environments).

The role of H$_{\rm 2}$ is of paramount importance throughout the formation of galaxies. 
Although debated and not yet fully understood, H$_{\rm 2}$ formation mainly occurs on dust grains in low-redshift and relatively metal-rich (Z~$\gtrsim 0.1$~Z$_{\odot}$) environments.  At higher redshift and in less-polluted regions,  H$_{\rm 2}$ formation usually results from the interaction between hydrogen (H) and electrons, and is mainly driven by the \ch{H-} catalysis -- i.e.  \ch{H + e- -> H- + $\gamma$} and \ch{H- + H -> H2 + e-} \citep[e.g.][for reviews]{Draine2003, Bromm2011}. 
Structure formation makes free electrons available across time, thus enhancing the \ch{H-} path to H$_{\rm 2}$ formation in cosmic environments \citep{Maio2022}.

From a numerical perspective, self-consistently accounting for the formation and evolution of H$_{\rm 2}$ is challenging: a complex set of rate equations has to be solved, and radiative-transfer computation has to be accounted for, as well. 
Moreover, the force and mass resolution of the majority of cosmological simulations falls short of the requirements for the GMCs (M$_{\rm GMC} \sim 10^5$~M$_{\odot}$) to be resolved. This implies that computations to retrieve H$_{\rm 2}$ have to be incorporated within sub-resolution models that also feature a description of the ISM structure.

Different methodologies can be exploited to compute H$_{\rm 2}$ and estimate the SFR from the molecular phase (see Section~\ref{discussion_H2} for details). Among them,  there are: $i)$ time-dependent, non-equilibrium rate equations; $ii)$ theoretical models that provide convenient approximations to different levels of refinement; $iii)$ observation-based, phenomenological prescriptions.  
Theoretical models belonging to the second aforementioned category 
\citep[e.g.][hereafter: KMT09]{Krumholz2008,Gnedin2009, KrumholzGnedin2011, GnedinKravtsov2011, Kr2009}
and phenomenological prescriptions falling into the third sub-sample 
\citep[e.g.][hereafter: BR06]{BR2006}
represent powerful tools to be exploited within sub-resolution models in cosmological simulations.

Most of the state-of-the-art cosmological hydrodynamical simulations compute the SFR from the available cold (T $\lesssim 10^4$~K) and dense (n$_{\rm H} \gtrsim 0.1 - 10$~cm$^{-3}$) gas, whose properties resemble those of the neutral H in the ISM. Among them, we can mention,  e.g.: Illustris \citep[][]{Vogelsberger2014}, Magneticum \citep[][]{Dolag2015}, NIHAO \citep[][]{Wang2015}, IllustrisTNG \citep[][]{Pillepich2018}, the Massive-Black \citep[][]{Khandai2015} and Fable \citep[][]{Henden2018} simulations, the Horizon-AGN suite \citep[][]{Dubois2016, Dubois2021} and the DIANOGA simulations \citep[][]{Bassini2020}. 
The Eagle simulation suite \citep[][]{Schaye2015} relies on a metallicity-dependent SF threshold \citep[][]{Schaye2004}: aiming at tying SF to the molecular gas, they exploit the CLOUDY radiative transfer code \citep[][]{Ferland1998} to determine the transition from the warm, atomic to the cold, neutral gas phase. 

Exceptions feature e.g.  the Mufasa and Simba simulations \citep[][]{Dave2017, Dave2019} and the FIRE suite \citep[][]{Hopkins2014, Hopkins2017},  all adopting the \citet{KrumholzGnedin2011} law to estimate the H$_{2}$ out of which stars are formed. 
Other examples of cosmological simulations targeting single haloes or smaller volumes with a sub-resolution treatment of H$_{\rm 2}$ include: \citet{Pelupessy2006, Feldmann2011, Christensen2012, Kuhlen2012, muppi2015, Tomassetti2015, Pallottini2019, Schaebe2020}. 

The latter aforementioned simulations have shown how crucial it is to convert only H$_{\rm 2}$ gas into stars. Indeed, besides achieving accuracy in the numerical modelling, getting rid of density and temperature ad-hoc thresholds, and moving closer to observational findings, this approach allows us to control the SFR by directly regulating the SF fuel \citep[e.g.][]{Kuhlen2012}. 
On the other hand, traditional prescriptions which compute the SFR from cold gas would rather prevent excessive SF by means of stellar feedback.  
An alternative, but less accurate approach to on-the-fly sub-resolution modelling of H$_{\rm 2}$ in simulations is represented by computing H$_{2}$ in post-processing \citep[e.g.][]{Popping2019}.

SF involves a hierarchy of physical processes and scales: the accretion of gas from the large-scale structure ($\sim$~Mpc), gas cooling to form the neutral phase (HI; $\sim$~kpc), the formation of GMCs ($\sim 10 - 100$~pc), H$_{2}$ fragmentation and accretion to make up clumps ($\sim 1$~pc) and cores ($\sim 0.1$~pc), and the subsequent contraction of cores to form stars ($\sim 10^{-8}$~pc) \citep{KennicuttEvans2012}.
This complexity has to be captured by (sub-resolution) models of structure formation, which also must account for those physical processes that interfere with SF. They include: feedback processes ensuing supernova (SN) explosions; black hole (BH) feedback possibly heating up the gas or removing it from sites of SF; turbulence and magnetic fields \citep{McKeeOstriker2007}.

In particular, massive stars can affect the reservoir of gas of a galaxy via feedback processes. These stem not only from the energy deposition and momentum injection following SN~explosions, but also from the ionizing effect that massive stars have before exploding \citep[usually referred to as early stellar feedback; e.g.][]{Stinson2013} and from stellar winds \citep[e.g.][]{Fierlinger2016, Fichtner2022}. 
In addition, the energy released by SNe~II is not expected to be constant across cosmic time nor in all the star-forming regions, and theoretical models predict that SNe exploding in almost pristine or weakly enriched environments provides the ISM with a larger amount of stellar feedback energy (see Section~\ref{discussion_EF} for references and details).  
As a consequence, cosmological simulations often adopt a non-constant value of the stellar feedback efficiency \citep[i.e. the fraction of the energy provided by each SN that is actually coupled to the surrounding gas as feedback energy; see e.g.][]{Schaye2015, Pillepich2018}. The aforementioned considerations motivated us to introduce a metallicity-dependent stellar feedback aimed at effectively capturing the outcome of SNe explosions in low-metallicity environments (i.e. low-metallicity stellar feedback, hereafter: LMF). 

In this paper, we aim at investigating the SF process in cosmological simulations of late-type galaxies across cosmic time. Our goal is twofold: $(a)$ to study how crucial it is to link the SFR to the reservoir of H$_{2}$ and how the latter quantity depends on the ISM properties; $(b)$ to investigate how the actual fuel for SF is sensitive to LMF throughout the star formation history (SFH) of forming structures. 
The main questions that we want to answer are the following: 
$(i)$ What is the impact of different prescriptions used to compute the H$_{2}$ content on galaxy formation and evolution? 
$(ii)$ Can final properties of simulated galaxies (e.g. their morphology) help in supporting or disfavouring different models? 
$(iii)$ How decisive is it to include a dependence on the ISM properties (e.g. gas metallicity) when modelling stellar feedback? 
To address the aforementioned questions, we perform cosmological hydrodynamical simulations targeting a Milky Way (MW)-size halo. 
We resort to the sub-resolution model MUPPI \citep[MUlti Phase Particle Integrator,][]{muppi2010, muppi2015} to describe the effect of physical processes happening at unresolved scales. This model features an advanced description of a multi-phase ISM and proved successful in zoomed-in simulations of late-type galaxies \citep[e.g.][]{Valentini2020, Granato2021, Giammaria2021}. In particular, within this sub-resolution model, we quantify the effect of assuming either the \citetalias[][]{BR2006} or the \citetalias[][]{Kr2009} model to calculate H$_{2}$.

The remainder of this paper proceeds as follows. Section~\ref{sec:gen_sims} introduces the new suite of cosmological simulations that we have carried out and describes how LMF and H$_{2}$-based SF are implemented.  Results are presented in Section~\ref{sec:results}, where we also perform some comparisons with observations, while in Section~\ref{sec:discussion} we discuss them within the current framework of state-of-the-art numerical works.  We draw conclusions in Section~\ref{sec:conclusions}.

\section{Cosmological simulations} 
\label{sec:gen_sims}

\subsection{Numerical set-up} 
\label{sec:MUPPI}

We carried out a set of cosmological hydrodynamical simulations to investigate how including LMF and modelling H$_{2}$ with different prescriptions affect the formation and evolution of late-type galaxies. 
We used the GADGET3 code, a non-public evolution of GADGET2 \citep{springel2005}, featuring the advanced formulation of smoothed particle hydrodynamics (SPH) introduced by \citet{beck2015} and coupled with MUPPI as in \citet{Valentini2017}. 
Our zoomed-in initial conditions (ICs) depict a dark matter (DM) halo 
with mass M$_{\rm halo, \, DM} \simeq 1.8 \times 10^{12}$~M$_{\odot}$ at redshift $z=0$.  
This halo was dubbed $AqC5$ when the ICs were originally introduced by \citet{Springel2008}. 
The zoomed-in region that our simulations target has been selected within a cosmological volume of 
$100 \, (h^{-1}$ Mpc$)^{3}$. 
We assume a $\Lambda$CDM cosmology, 
with $\Omega_{\rm m}=0.25$, 
$\Omega_{\rm \Lambda}=0.75$, 
$\Omega_{\rm baryon}=0.04$, 
$\sigma_8 = 0.9$, $n_s=1$, 
and $H_{\rm 0}=100 \,h$ km s$^{-1}$ Mpc$^{-1}=73$ km s$^{-1}$ Mpc$^{-1}$. 
Mass and force resolution are as follows: 
DM particles have a mass of $1.6 \times 10^6 \, h^{-1}$~M$_{\odot}$, 
the initial mass of gas particles is $m_{\rm gas, in} = 3.0 \times 10^5 \, h^{-1}$~M$_{\odot}$. 
As for the computation of the gravitational force, we adopt a Plummer-equivalent softening length of 
$\varepsilon_{\rm Pl} = 325 \, h^{-1}$~pc. This is kept constant, before in comoving units down to $z=6$ and then in physical units. 
The aforementioned simulation set-up has been already used in e.g. \citet[][]{Valentini2017, Valentini2020} and \citet[][]{Granato2021}.

\subsection{The sub-resolution model} 
\label{sec:MUPPI}

Our simulations adopt the sub-resolution model MUPPI 
\citep[][]{muppi2010, muppi2015, Valentini2017, Valentini2019, Valentini2020} 
to account for unresolved physical processes. 
A comprehensive and up-to-date description of MUPPI as used in the simulations presented in this paper can be found in Sections~$2$~and~$3$ of \citet{Valentini2020}. Reference values of all the relevant parameters of MUPPI (apart from those explicitely mentioned below) are those adopted in the aforementioned paper. Here we only recall the most relevant features of the model and only focus on the numerical description of newly introduced or updated physical processes that are relevant for the present investigation.

MUPPI describes a multiphase ISM. Gas particles whose density rises above a density threshold ($n_{\rm H, \, thres}=0.01$~cm$^{-3}$) and whose temperature falls below a temperature threshold ($T_{\rm thresh}=5 \times 10^4$~K) enter the so-called multi-phase stage. 
These multi-phase particles feature a hot and a cold gas phase in pressure equilibrium, besides a possible, virtual stellar component.  The three phases within each multi-phase particle exchange mass and energy, these flows being defined by a set of differential equations.  
MUPPI includes the following physical processes: radiative cooling and gas evaporation due to the destruction of molecular clouds; 
a stochastic model for star formation \citep[following][]{SpringelHernquist2003} based on the availability of molecular gas (see below); 
thermal and kinetic stellar feedback \citep{muppi2015, Valentini2017};
a model for angular-momentum dependent gas accretion onto BHs and isotropic, thermal AGN (active galactic nucleus) feedback \citep{Valentini2020}.
Our simulations also feature a model of chemical evolution \citep{tornatore2007}. Stellar particles represent simple stellar populations (SSPs). By assuming an initial mass function (IMF; in this work we assume \citet{kroupa93}, see below), stellar lifetimes, and stellar yields \citep[see][for details]{Valentini2019, Valentini2020}, the number of aging and exploding stars is computed,  along with the amount of metals returned to the ISM.  As for the metallicity-dependent radiative cooling,  each particle contributes to the cooling rate \citep{wiersma2009}. 
The presence of a spatially uniform, time-dependent ionizing cosmic background 
\citep{HaardtMadau2001} is accounted for when evaluating cooling rates. 

Within the H$_{2}$-based model for star formation of MUPPI, a fraction $f_{\rm H_{2}}$ of the cold ($T_{\rm c} = 300$~K) gas mass within each multi-phase particle, $M_{\rm c}$, is expected to be in the molecular phase and converted into stars with a constant efficiency. We adopt $f_{\ast} = 0.02$ as efficiency of star formation per dynamical time \citep[][]{Kennicutt1998, KrumholzTan2007, Tacconi2020}.
The particle SFR, $\dot{M}_{\rm sf}$, thus reads: 
\begin{equation}
\centering
\dot{M}_{\rm sf} = f_{\ast} \, \frac{f_{\rm H_{2}} \, M_{\rm c}}{t_{\rm dyn, c}} \,,
\label{eq:sfr}
\end{equation}
\noindent
where $t_{\rm dyn, c} = [ 3 \, \pi / (32 \, G \, \rho_{\rm c})]^{1/2}$ is the dynamical time of the cold phase and $\rho_{\rm c}$ the density of the cold gas.  Section~\ref{sec:H2} addresses how we computed $f_{\rm H_{2}}$. 

At variance with the simulations in \citet{Valentini2020}, we updated the computation of the timescale exploited to retrieve the SFR. Now, we assume that the depletion timescale for SF,  $t_{\rm dyn, c}$, is computed by considering the cold gas density at the time when the SFR of each multiphase particle  is estimated\footnote{The MUPPI model 
	integrates a system of differential equations. 
	A Runge-Kutta algorithm is adopted to integrate equations within each SPH time-step, 
	and a multi-phase stage can last up to several SPH time-steps 
	\citep[see][for details]{muppi2010, muppi2015, Valentini2020}. }.  
This is dissimilar from previous simulations adopting MUPPI, where the $t_{\rm dyn, c}$ was computed as soon as the cold gas in the considered particle exceeded the $95$ per cent of the particle mass; once computed, the value was assumed to remain constant for the entire multi-phase stage of the particle \citep[see discussion in Section~2.2 of][for details]{muppi2010}. The origin for this change stems from the need of adopting a depletion timescale for SF that is more sensitive to the way in which the thermodynamical properties of the cold ISM vary. The updated computation of $t_{\rm dyn, c}$ makes it on average shorter than previously estimated, thus allowing a more bursty SF especially at high redshift. A detailed investigation of this effect will be the topic of an upcoming paper (Valentini et al. 2022, in preparation).

Due to this modification, we also re-tuned three parameters with respect to \citet{Valentini2020}\footnote{Previously adopted values were: 
$P_{\rm kin} = 0.05$, 
$t_{\rm wind} = 15 - t_{\rm dyn, c \,\, [Myr]}$~Myr, 
and $f_{\rm fb, kin} = 0.26$.}. 
The probability for a gas particle to become a wind particle and sample galactic outflows is set to $P_{\rm kin} = 0.03$, 
while the maximum lifetime of wind particles is $t_{\rm wind} = 45 - t_{\rm dyn, c \,\, [Myr]}$~Myr. 
Moreover, the kinetic stellar feedback efficiency that we adopt is $f_{\rm fb, kin} = 0.1$: decreasing this efficiency turned out to be key to control the excessive gas fall-back and high SFR at low redshift \citep[limitations of our previous simulations are discussed in][]{Valentini2019, Granato2021}, while still enabling the formation of an extended stellar disc.

\subsubsection{Estimating the molecular gas} 
\label{sec:H2}

We adopt the following two different prescriptions to estimate $f_{\rm H_{2}}$.
\begin{description}

\item[$\bullet$ \bf{Blitz \& Rosolowsky}] {\sl (BR)} - From observations of late-type galaxies in the local Universe including our MW,  \citetalias{BR2006} found that the interstellar gas pressure determines the ratio between the atomic and molecular gas surface densities. They showed  that the molecular gas fraction $f_{\rm H_{2}}$ is proportional to the pressure $P$ of the gas in the galaxy disc, i.e. $f_{\rm H_{2}} \propto P^{\, \alpha}$, with the best-fit parameter $\alpha = 0.92 \pm 0.07$. 
By approximating $\alpha \simeq 1$,  their phenomenological prescription can be cast as \citep[see also][]{muppi2010}: 
\begin{equation}
\centering
f_{\rm H_{2}} \:=\: f_{\rm H_{2}}(n_{\rm c}) \:=\: \frac{1}{1+P_0/P} \,\,, 
\label{eq:f_mol}
\end{equation}
\noindent
where $n_{\rm c}$ is the cold gas number density, and the pressure of the ISM $P_0$ at which $f_{\rm H_{2}}=0.5$ is a parameter derived from observations.  
Specifically, the galaxy sample of local spiral galaxies of \citetalias{BR2006} predicts that $P_0 / {\text {k}}_{\rm B}$ spans the range ($0.4$~--~$7.1$)~$\times 10^4$~K~cm$^{-3}$,  ${\text {k}}_{\rm B}$ being the Boltzmann constant. 
We adopt $P_0 / {\text {k}}_{\rm B}= 2 \times 10^4$~K~cm$^{-3}$.  
In the original formalism by \citetalias{BR2006}, $P$ is the hydrostatic pressure of the gas in the galactic disc; within our model, $P$ is the hydrodynamic pressure of the gas particle. 
Equation~(\ref{eq:f_mol}) can be exploited to retrieve the effective density threshold for SF in MUPPI when the BR model is adopted, and to appreciate how much it exceeds $n_{\rm H, \, thres}=0.01$ cm$^{-3}$, which is the density threshold for a particle to become multiphase. Considering $n_{\rm thresh, sf}$ as the number density of the cold gas for which $f_{\rm H_{2}}=0.5$ and plugging in the assumed values for $P_0$ and $T_{\rm c}$,  the effective number density threshold for SF is $n_{\rm thresh, sf}\simeq 66.7$~cm$^{-3}$. This also explains the dependency of $f_{\rm H_{2}}$ on $n_{\rm c}$.\\

\item[$\bullet$ \bf{Krumholz, McKee \& Tumlinson}] {\sl (Kr)} - In a series of papers, \citet{Krumholz2008, Krumholz2009ApJ693, Kr2009} showed that $f_{\rm H_{2}}$ in a galaxy is primarily controlled by gas surface density, with gas metallicity playing a supporting role at determining it, and with the strength of the interstellar radiation field contributing little to nothing. 
Their theoretical model predicts $f_{\rm H_{2}}$ by imposing that $H_{2}$ formation on dust grains counterbalances $H_{2}$ destruction by Lyman-Werner photons. Their formulation relies on the semi-analytical model by \citet{Wolfire2003}, calibrated on the ISM properties of the MW. According to the \citetalias{Kr2009} model $f_{\rm H_{2}}$ reads:
\begin{equation}
\centering
f_{\rm H_{2}} \:=\: f_{\rm H_{2}}(\Sigma, Z') \:=\: 1 - \biggl( 1 + \biggl( \frac{3}{4} \frac{s}{1+\delta} \biggr)^{- 5} \biggr)^{- 1/5} \,\,, 
\label{eq:f_mol_kr}
\end{equation}
where:
$ s \:=\: \text{ln}(1 + 0.6 \chi) / ( 0.04 \, \Sigma \, Z') $, 
$ \,\, \chi \:=\: 0.77 \, (1 + 3.1 Z'^{0.365}) $, 
$ \,\, \delta \:=\: 0.0712 \, ( 0.1 \, s^{-1} + 0.675)^{-2.8} $, and 
$ Z' = Z/Z_{\odot}$ is the gas metallicity normalised to the Solar metallicity.  Throughout the paper, we refer to the metallicity $Z$ as the total mass of all the elements heavier than Helium that we track for each particle in our simulations,  divided by the gas mass. We adopt the present-day value $Z_{\odot}=0.01524$ \citep{Caffau2011}.  
In equation~(\ref{eq:f_mol_kr}), $\Sigma$ is the gas surface density expressed in units of M$_{\odot}$~pc$^{-2}$.  In the formalism by \citetalias{Kr2009},  the $\Sigma$ entering equation~(\ref{eq:f_mol_kr}) is the surface density of a representative, $\sim 100$~pc-sized atomic-molecular gas complex.
Should the sampled patch of the ISM be significantly different in size (namely, if gas properties are averaged over larger regions in low-resolution simulations), the estimated $\Sigma$ should be boosted by a clumping factor (of order of $\sim$ a few) before being plugged in equation~(\ref{eq:f_mol_kr}). Multi-phase particles in our simulations sample the densest regions of the galaxy where the size of the particle volume approaches the formal force resolution $\varepsilon_{\rm Pl}$, and their mass is on average made up of cold and molecular gas by $>90\%$; nonetheless, without such a clumping factor we would estimate gas surface densities which are too low for the \citetalias{Kr2009} model to correctly predict $f_{\rm H_{2}}$ via equation~(\ref{eq:f_mol_kr}).  
To estimate the clumping factor $C$, we exploited results of \citet[][see their Section~2.3 and Table~1]{Dave2016}. The clumping factor that we adopt can be cast as: $C= 0.1 \cdot \rho_{\rm max, Dave} / \rho_{\rm max, fid}$.  Here, $\rho_{\rm max, Dave}$ is the density estimated from the mass and force resolution of \citet[mass and softening length of gas particles listed in their Table~1]{Dave2016}, while $\rho_{\rm max, fid}$ is the same quantity that we retrieve from our simulations at the fiducial resolution considered in this work. 
Specifically, $\rho_{\rm max, fid} = m_{\rm gas, in} / \varepsilon_{\rm Pl}^{3} = 4.66 \times 10^6$~M$_{\odot}$~kpc$^{-3}$.  
By adopting the normalization factor $0.1$, we find that boosting $\Sigma$ by about $4$ is needed at our fiducial resolution. The surface density $\Sigma$ multiplied by the clumping factor $C$ enters equation~(\ref{eq:f_mol_kr}). 
We evaluate $\Sigma$ on a particle base via the Sobolev approximation, i.e.  $\Sigma \propto \rho ^2 / \, | \nabla \rho |$ \citep[as suggested by][]{KrumholzGnedin2011}.  
A metallicity floor ($Z_{\rm floor} = 0.05 \, Z_{\odot}$) is imposed when computing $f_{\rm H_{2}}$, the \citetalias{Kr2009} model being not valid below that threshold: we assume $Z_{\rm floor}$ to evaluate $f_{\rm H_{2}}$ for star-forming particles having $Z<Z_{\rm floor}$.
\end{description}

\subsubsection{Stellar feedback from low-enrichment environments} 
\label{sec:EF}

Aiming at mimicking the larger amount of feedback energy that SNe from Population III stars are expected to inject in a pristine ISM (see Section~\ref{discussion_EF} for details), we assume that star-forming particles with $Z < 0.05 \, Z_{\odot}$ produce a stronger stellar feedback. The larger amount of feedback energy available from low-metallicity star-forming particles is distributed as thermal, kinetic, and local contributions without modifying the stellar feedback scheme described in \citet{Valentini2020}, apart from the value of $f_{\rm fb, kin}$ (as discussed above). 
Feedback energy provided by each SN~II from star-forming particles with $Z < 0.05 \, Z_{\odot}$ is boosted with reference to the fiducial $E_{\rm SN} = 10^{51}$~erg by a factor that is a parameter of the model (see Section~\ref{sec:sims} and Appendix~\ref{AppA}).

In all the simulations presented in this work, we adopt the \citet{kroupa93} IMF to estimate, e.g., the number of stars exploding as SNe. 
We do not modify the IMF for low-metallicity ($Z < 0.05 \, Z_{\odot}$), star-forming particles, although Population III stars are likely expected to be characterized by a more top-heavy IMF \citep{Bromm2004}. 
This effective description mainly aims at understanding the impact of LMF on the suppression of SF in those regions where first stars have formed and just exploded.
We postpone a more detailed and accurate description of LMF to a forthcoming paper.

\subsection{The suite of simulated galaxies} 
\label{sec:sims}

Table~\ref{tab:sims} introduces our simulations and lists their key features. The simulation label encodes the model adopted to estimate H$_{2}$, either {\sl {Kr}} \citepalias{Kr2009} or {\sl {BR}} \citepalias{BR2006}, and whether LMF is accounted for or not ({\sl {noLMF}}). The suffix \_fid highlights our fiducial model. 

When LMF is included with a boosting factor for the stellar feedback energy different from the reference one (i.e. $10$x), the label of the run also contains the adopted fudge factor (e.g. a factor $5$ or $20$ boost is indicated with \_$5$x or \_$20$x, respectively). 
The suffix \_fmf highlights that the fraction of the molecular gas is fixed to $0.95$ for low-metallicity (Z~$<0.05\ ~$Z$_{\odot}$) star-forming particles. We consider two additional simulations for reference: \textcolor{Gray}{\bf Kr\_noAGN} where AGN feedback is not accounted for and \textcolor{Black}{\bf Kr\_noBH} where not even BHs are included (see Appendix~\ref{AppB}). 

Besides the runs listed in Table~\ref{tab:sims},  other simulated galaxies will be presented in Appendices~\ref{AppC} and~\ref{AppD}.  In particular, in Appendix~\ref{AppC} we will consider also ICs different from the ones introduced in Section~\ref{sec:MUPPI}, to test our fiducial model in different haloes.

\begin{table}
\centering
\begin{minipage}{85mm}
\newcommand{\cmark}{\ding{51}}
\newcommand{\xmark}{\ding{55}}
\caption[]{Relevant features of the simulation suite. 
{\sl {Column~1:}} simulation label color-coded as in the figures. 
{\sl {Column~2:}} model adopted to estimate H$_{2}$, either 
Kr \citepalias{Kr2009} or BR \citepalias{BR2006}.
{\sl {Column~3:}} whether stellar feedback from low-metallicity environments (LMF) is implemented (\cmark) or not (\xmark). 
{\sl {Column~4:}} factor by which the stellar feedback energy supplied by 
low-metallicity (Z~$<0.05\ ~$Z$_{\odot}$) star-forming particles is boosted with respect to 
the original model without LMF. } 
\renewcommand\tabcolsep{3.5mm}
\begin{tabular}{@{}lccc@{}}
\hline
Simulation  &  Model            &  Low-Metallicity       &  LMF Energy       \\ 
                    &  for H$_{2}$  &   stellar Feedback (LMF)  &    Booster         \\ 
\hline
\hline
\textcolor{Blue}{\bf Kr\_~fid}   &    Kr    &    \cmark   &       $10$x           \\ 
\hline
\textcolor{Cerulean}{\bf Kr\_noLMF}                    &   Kr     &   \xmark    &        \xmark          \\
\hline
\textcolor{YellowOrange}{\bf BR}           &    BR   &    \cmark   &       $10$x           \\  
\hline
\textcolor{Red}{\bf BR\_noLMF}                   &   BR     &   \xmark    &        \xmark          \\ 
\hline
\textcolor{Plum}{\bf Kr\_5x}    &    Kr    &    \cmark   &       $5$x           \\ 
\hline
\textcolor{WildStrawberry}{\bf Kr\_20x}   &    Kr    &    \cmark   &       $20$x           \\ 
\hline
\textcolor{LimeGreen}{\bf Kr\_fmf}   &    $\;\;\;$Kr~\footnote{This model also assumes that the molecular gas fraction is fixed to $0.95$ for low-metallicity (Z~$<0.05\ ~$Z$_{\odot}$) star-forming particles.}   &    \cmark   &       $10$x           \\ 
\hline
\textcolor{Black}{\bf Kr\_noBH}   &    Kr    &    \cmark   &       $10$x           \\ 
\hline
\textcolor{Gray}{\bf Kr\_noAGN}   &    Kr    &    \cmark   &       $10$x           \\ 
\hline
\hline
\end{tabular}
\label{tab:sims}
\end{minipage}
\end{table}

\section{Results} 
\label{sec:results}

\begin{figure*}
\newcommand{\captionfonts}{\small}
\begin{minipage}{\linewidth}
\centering
\vspace{-2.ex}
\includegraphics[trim=0.5cm 2.3cm 10.cm 1.5cm, clip, width=.68\textwidth]{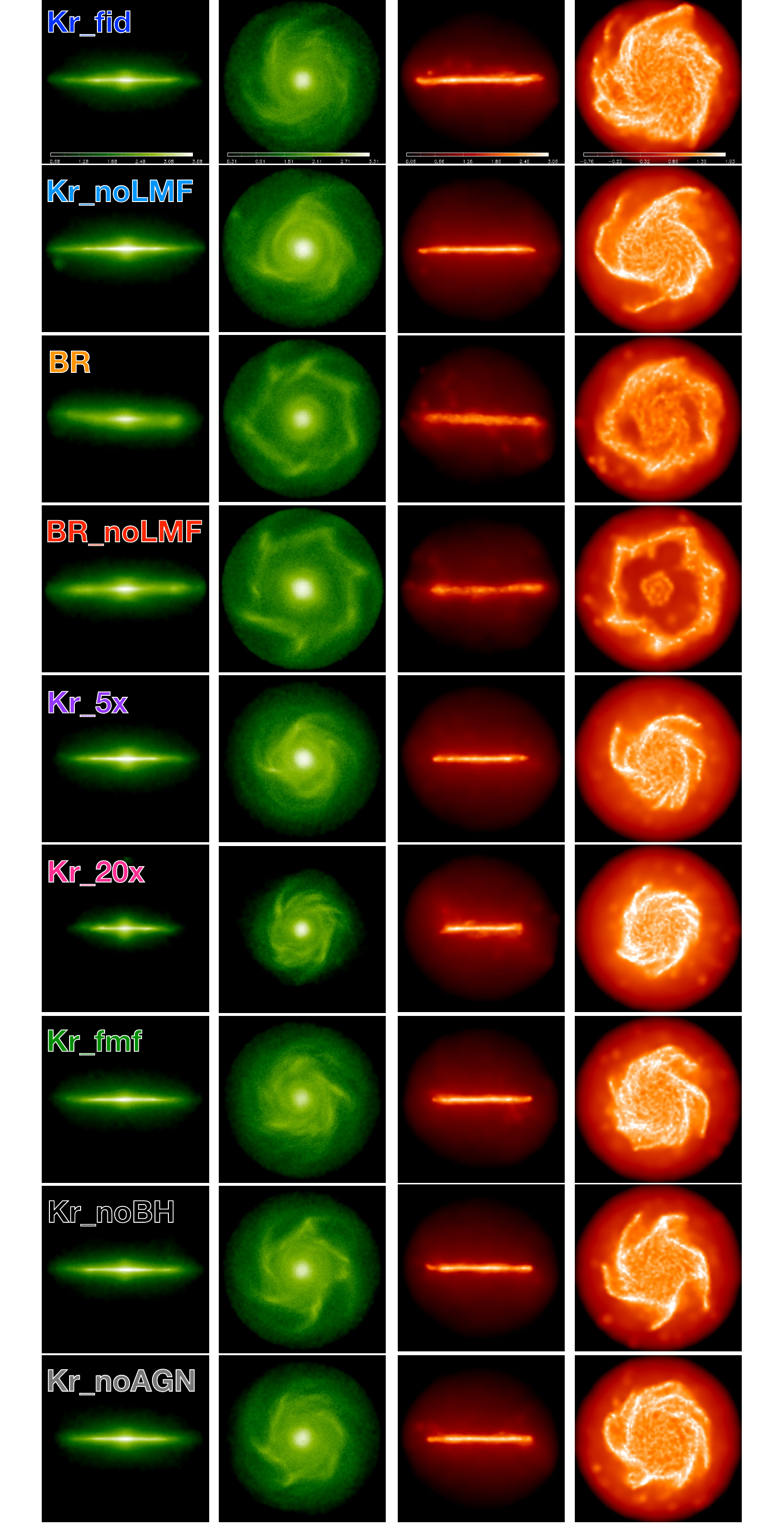} 
\end{minipage} 
\vspace{-2.ex}
\caption{Projected stellar (first and second columns) and gas (third and forth columns) density maps for the simulated galaxies listed in Table~\ref{tab:sims} and written on the first-column panels, at~$z = 0$. 
First and third columns show edge-on galaxies, second and forth ones depict face-on maps. Each box is $50$~kpc a side.  Colour bars encode the logarithm of projected densities (M$_{\odot}$ pc$^{-2}$) and are common to panels in each column. }
\label{StellarGasMaps} 
\end{figure*}

Figure~\ref{StellarGasMaps} introduces the full suite of simulated galaxies and shows projected stellar and gas density maps, at $z = 0$. The z-axis of the galaxy reference system is aligned with the angular momentum of star and (cold and multiphase) gas particles within $8$~kpc from the minimum 
of the gravitational potential. 
The centre of mass of the aforementioned particles is assumed to be the origin of the galaxy reference system. Throughout this paper, our analysis focuses on star and gas particles located within $0.1$~r$_{\rm vir}$, i.e. the galactic radius\footnote{We define the galactic 
	radius as one tenth of the virial radius, i.e. ~r$_{\rm gal}=0.1$~r$_{\rm vir}$. 
  	The radius~r$_{\rm gal}$ is chosen to select the region of the 
  	computational domain where the central galaxy resides. 
  	We consider virial quantities as those computed in a sphere that encompasses 
	an overdensity of 200 times the critical density at the considered redshift and 
	that is centred on the minimum of the gravitational potential of the halo.}, unless otherwise specified.  Virial radii of these galaxies are~$\approx 240$~kpc (see Table~\ref{tab:GasMasses}).  

By visual inspection,  Figure~\ref{StellarGasMaps} shows how different numerical prescriptions and the details entering the modelling of physical processes affect the morphology of simulated galaxies. Indeed, while all the galaxies have a clear disc-dominated morphology, the prominence of the stellar bulge, the spatial extension of the disc, and the details of the spiral pattern represent a distinctive signature of the considered model. As a consequence, specific properties of simulated galaxies can help in supporting or discarding features of the underlying galaxy formation model, especially when predictions from simulations are compared with observations. 

In what follows, we focus our analysis on a reference sub-sample of models, namely: 
\textcolor{Blue}{\bf Kr\_~fid},  \textcolor{Cerulean}{\bf Kr\_noLMF}, 
\textcolor{YellowOrange}{\bf BR}, and \textcolor{Red}{\bf BR\_noLMF}. 
The remaining simulated galaxies will be further considered in Appendices~\ref{AppA},~\ref{AppB}, ~\ref{AppC} and~\ref{AppD}.

\subsection{The build-up of the galaxy disc} 
\label{subsec:SFH}

\begin{figure}
\newcommand{\captionfonts}{\small}
\begin{minipage}{\linewidth}
\centering
\includegraphics[trim=0.2cm 0.cm 0.2cm 0.2cm, clip, width=1.\textwidth]{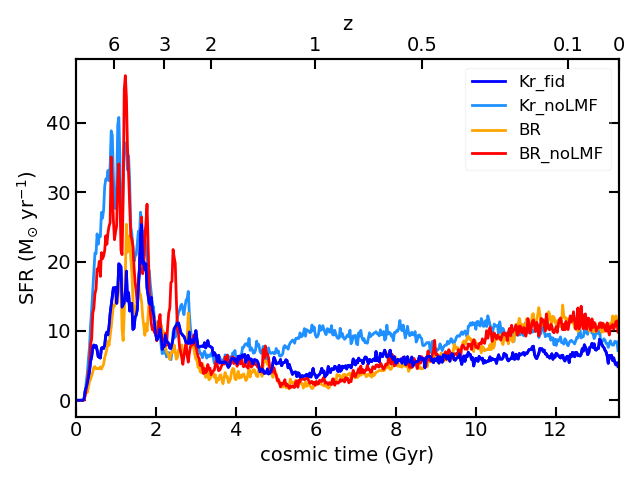} 
\end{minipage} 
\caption{Star formation history of galaxy models 
\textcolor{Blue}{\bf Kr\_~fid},  \textcolor{Cerulean}{\bf Kr\_noLMF}, 
\textcolor{YellowOrange}{\bf BR}, and \textcolor{Red}{\bf BR\_noLMF}. 
Including LMF reduces the high-z SF burst.}
\label{SFH} 
\end{figure}

We analyse the SFH of simulated galaxies to assess the effect of LMF and different prescriptions for SF across cosmic time. 
Figure~\ref{SFH} presents the evolution of the SFR for the reference set of four galaxy models. SFRs are retrieved from the stellar ages of star particles within~r$_{\rm gal}$ at $z=0$.

LMF leaves a clear imprint at high-redshift, significantly reducing the SF burst (at $z \gtrsim 3$) for \textcolor{Blue}{\bf Kr\_~fid} and \textcolor{YellowOrange}{\bf BR}. Indeed, the additional stellar feedback energy injection is able to suppress and delay SF at high-z. Gas heated up by LMF at high-z remains hotter and less prone to SF also later on ($2.5 \gtrsim z \gtrsim 1.5$), when the two aforementioned models still have lower SFRs with respect to \textcolor{Cerulean}{\bf Kr\_noLMF} and \textcolor{Red}{\bf BR\_noLMF}. The milder high-z SF burst experienced by \textcolor{Blue}{\bf Kr\_~fid} with respect to \textcolor{Cerulean}{\bf Kr\_noLMF} also results in a reduced gas fall-back (following gas ejection by SN explosions) on the forming galaxy below $z \lesssim 2$, thus contributing to maintain lower SFRs. This effect is not visible when comparing models \textcolor{YellowOrange}{\bf BR} and \textcolor{Red}{\bf BR\_noLMF}. 
At low z ($z \lesssim 1$), the H$_{2}$ modelling mainly drives the SFR: BR galaxies have a larger SFR. At $z=0$, the SFR of \textcolor{Blue}{\bf Kr\_~fid} is smaller by $\gtrsim 2x$ than \textcolor{YellowOrange}{\bf BR}, and the lower SFR of Kr models is independent of whether LMF is adopted.

\begin{figure}
\newcommand{\captionfonts}{\small}
\begin{minipage}{\linewidth}
\centering
\includegraphics[trim=0.2cm 0.cm 0.2cm 0.2cm, clip, width=1.\textwidth]{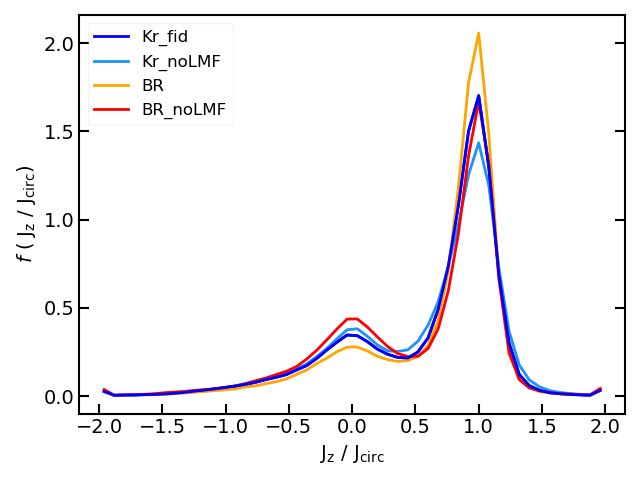} 
\end{minipage} 
\caption{Kinematic decomposition of the stellar mass within $0.1$~r$_{\rm vir}$
of galaxies \textcolor{Blue}{\bf Kr\_~fid},  \textcolor{Cerulean}{\bf Kr\_noLMF}, 
\textcolor{YellowOrange}{\bf BR}, and \textcolor{Red}{\bf BR\_noLMF}.  
The ratio of specific angular momenta J$_{\rm z}$/J$_{\rm circ}$ tells apart the stellar mass 
in the bulge (J$_{\rm z}$/J$_{\rm circ} \simeq 0$) 
from that in the galaxy disc (J$_{\rm z}$/J$_{\rm circ} \simeq 1$). The y axis shows the fraction of the total stellar mass in each bin of J$_{\rm z}$/J$_{\rm circ}$, divided by the bin width, for each galaxy model.}
\label{jcirc} 
\end{figure}

\begin{figure}
\newcommand{\captionfonts}{\small}
\begin{minipage}{\linewidth}
\centering
\includegraphics[trim=1.2cm 0.3cm 0.2cm 1.5cm, clip, width=1.\textwidth]{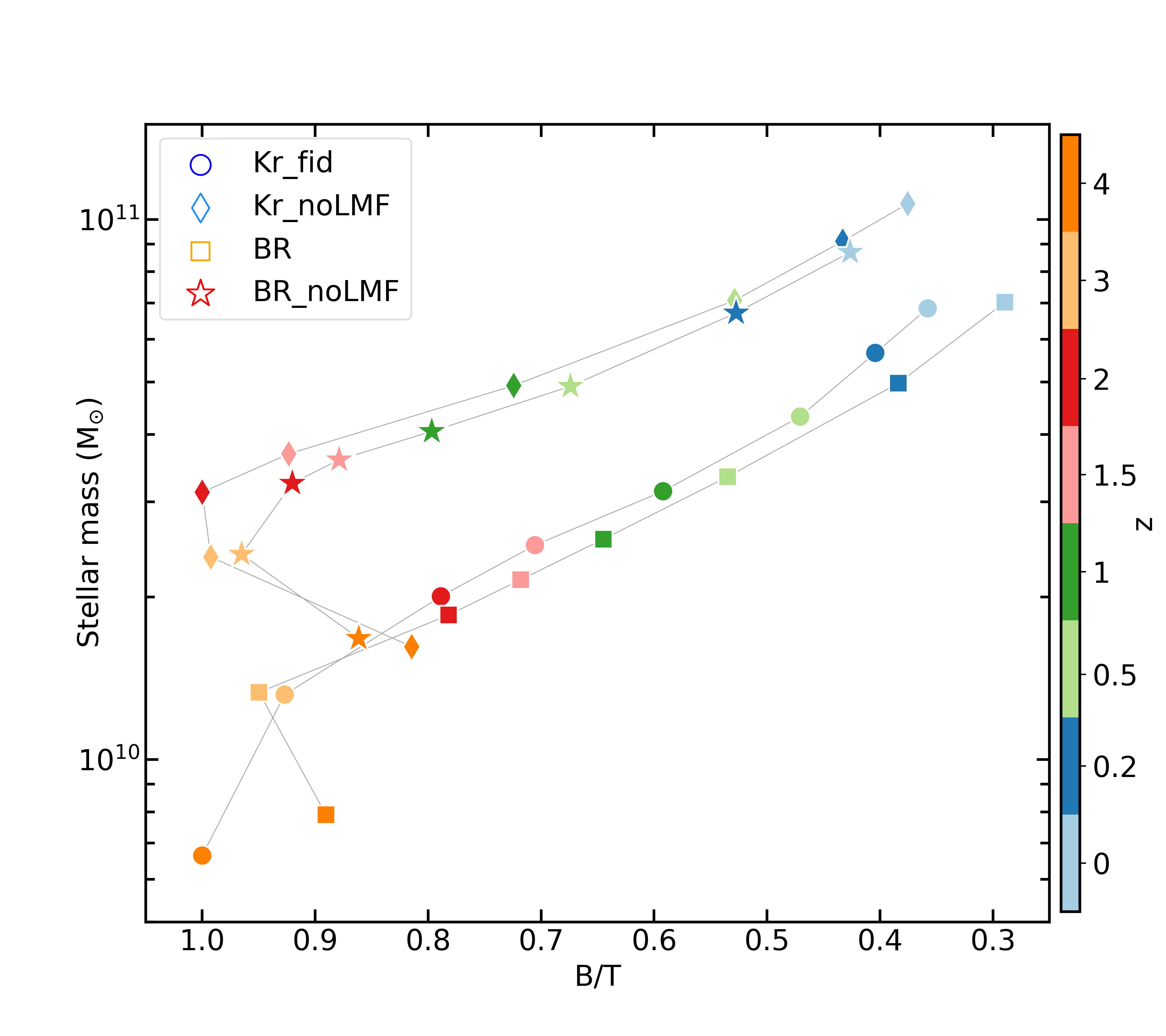} 
\end{minipage} 
\caption{Evolution of galaxy stellar mass as a function of the bulge-over-total mass ratio B/T, 
for models \textcolor{Blue}{\bf Kr\_~fid},  \textcolor{Cerulean}{\bf Kr\_noLMF}, 
\textcolor{YellowOrange}{\bf BR}, and \textcolor{Red}{\bf BR\_noLMF}.  
Each model is characterised by a different symbol. 
Symbol colours encode the redshift, as detailed in the color bar.
Galaxies usually become disc-dominated (B/T~$\lesssim 0.5$) below $z \sim 0.2$.}
\label{BsuT} 
\end{figure}

\begin{table*}
\centering
\begin{minipage}{178mm}
\caption[]{Relevant gas and stellar masses for different simulations ({\sl {Column~1}}) at $z=0$. 
{\sl {Column~2:}} virial radius of the simulated galaxy. 
{\sl {Columns~3~and~4:}} gas mass within~r$_{\rm vir}$ and~r$_{\rm gal}$, respectively. 
{\sl {Column~5:}} cold ($T_{\rm cold}=300$~K) gas mass within~r$_{\rm gal}$, including molecular gas.
{\sl {Column~6:}} neutral gas mass within~r$_{\rm gal}$, i.e. the hydrogen mass in the cold phase minus M$_{\rm mol}$.
{\sl {Column~7:}} molecular gas mass within~r$_{\rm gal}$.
{\sl {Column~8:}} average molecular to cold gas mass ratio, within~r$_{\rm gal}$.
{\sl {Column~9:}} galaxy stellar mass, within~r$_{\rm gal}$.
{\sl {Column~10:}} bulge-over-total stellar mass ratio.
{\sl {Column~11:}} galaxy effective radius with standard deviation.
Note that all the $\text{M}_{\rm cold}$ and $\text{M}_{\rm mol}$ within~r$_{\rm vir}$ are located within~r$_{\rm gal}$.} 
\renewcommand\tabcolsep{3.2mm}
\begin{tabular}{@{}lccccccccccc@{}}
\hline
Simulation  &  $\text{r}_{\rm vir}$  &
\multicolumn{2}{c}{$\text{M}_{\rm gas}$  } &   
$\text{M}_{\rm cold}$  &   
$\text{M}_{\rm HI}$  &  
$\text{M}_{\rm mol}$   &  $\text{M}_{\rm mol}/\text{M}_{\rm cold}$  &  $\text{M}_{\ast}$  &  B/T   & 
$\text{r}_{\rm s}$   \\ 
              & (kpc) &   \multicolumn{2}{c}{($10^{11}$~$\text{M}_{\odot}$) }  &  ($10^{10}$~$\text{M}_{\odot}$)   &  ($10^{10}$~$\text{M}_{\odot}$)   &  ($10^9$~$\text{M}_{\odot}$) & ($10^{-1}$)  &  ($10^{10}$~$\text{M}_{\odot}$)  & & (kpc)  \\
              & &  r <  r$_{\rm vir}$   &  < r$_{\rm gal}$   &  < r$_{\rm gal}$   &  < r$_{\rm gal}$    &  < r$_{\rm gal}$  & & < r$_{\rm gal}$  & \\ 
\hline
\hline
\textcolor{Blue}{\bf Kr\_~fid}      & $240.8$ &  $1.18$  &  $0.24$  &   $1.68$ &   $1.07$ &    $1.72$ & $1.02$ & $6.84$ & $0.36$  & $4.1 \pm 0.1$ \\  
\hline
\textcolor{Cerulean}{\bf Kr\_noLMF}                & $243.8$ &  $1.37$  &  $0.21$ &  $1.26$  &  $0.73$  &   $1.87$ & $1.48$ & $10.69$ & $0.38$  & $4.1 \pm 0.1$ \\  
\hline
\textcolor{YellowOrange}{\bf BR}                & $240.8$ &  $1.23$  &  $0.14$  &   $0.80$  &   $0.31$ &   $2.80$  & $3.50$ & $7.02$ & $0.29$  & $4.4 \pm 0.2$ \\  
\hline
\textcolor{Red}{\bf BR\_noLMF}      & $242.7$ &  $1.42$  &  $0.12$  &    $0.66$  &    $0.23$ &   $2.45$  & $3.71$ & $8.69$  & $0.43$  & $4.3 \pm 0.2$ \\  
\hline
\hline
\end{tabular}
\label{tab:GasMasses}
\end{minipage}
\end{table*}

\begin{figure*}
\newcommand{\captionfonts}{\small}
\begin{minipage}{\linewidth}
\centering
\includegraphics[trim=0.5cm 0.cm 3.cm 0.5cm, clip, width=1.\textwidth]{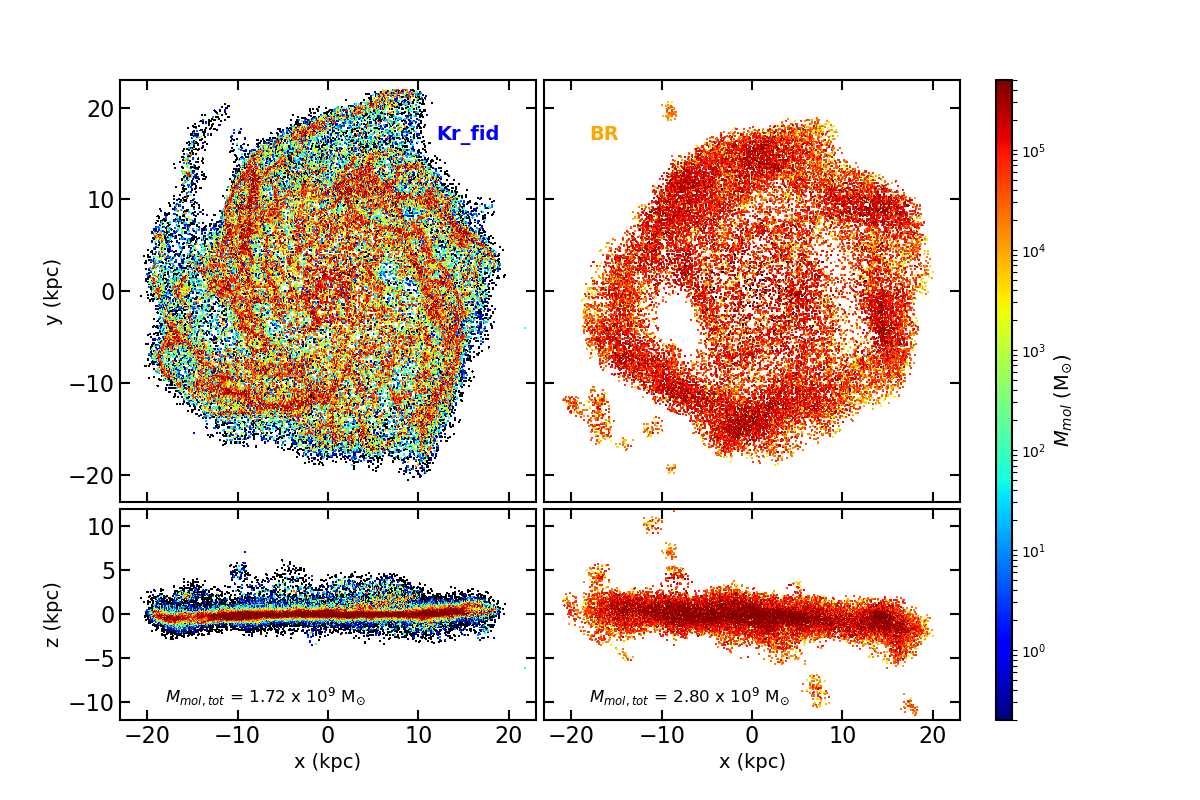} 
\end{minipage} 
\caption{Face-on ({\sl top panels}) and edge-on ({\sl bottom panels}) binned distributions 
of all the star-forming particles containing molecular gas and located within~$0.1$~r$_{\rm vir}$ 
of the simulated galaxies \textcolor{Blue}{\bf Kr\_~fid} ({\sl left panels}) 
and \textcolor{YellowOrange}{\bf BR} ({\sl right panels}).  
Plots are shown at $z = 0$; 
the colour encodes the total H$_{2}$ mass in the bin.}
\label{MmolMaps} 
\end{figure*}

\begin{figure*}
\newcommand{\captionfonts}{\small}
\begin{minipage}{\linewidth}
\centering
\includegraphics[trim=0.4cm 0.cm 0.4cm 0.cm, clip, width=1.\textwidth]{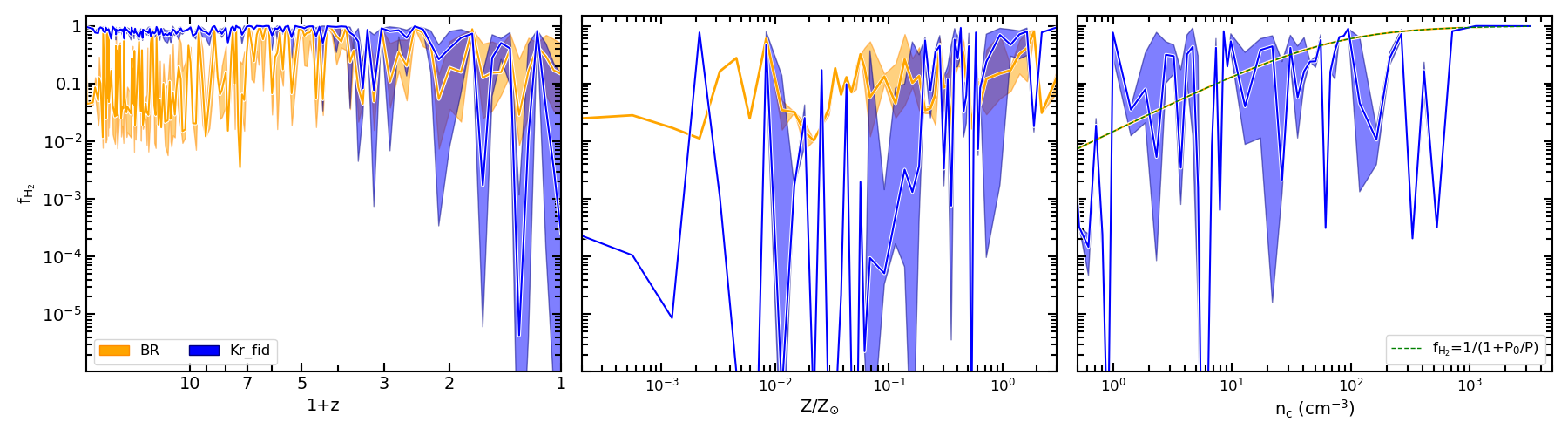} 
\end{minipage} 
\caption{Dependency of the molecular gas fraction on redshift ({\sl left panel}), 
gas metallicity ({\sl centre}),  and cold gas number density ({\sl right panel})
for models \textcolor{Blue}{\bf Kr\_~fid}  
and \textcolor{YellowOrange}{\bf BR}.  
Solid lines describe the median evolution 
of (a randomly-selected sub-sample of) star-forming particles in the considered bin, shaded envelopes show the region between the 16th and 84th percentile of the particle distribution in each bin. In the central and right panels we consider the full sub-sample of star-forming particles across time.}
\label{fmolMaps} 
\end{figure*}

\begin{figure*}
\newcommand{\captionfonts}{\small}
\begin{minipage}{\linewidth}
\centering
\includegraphics[trim=3.2cm 0.cm 3.8cm 0.5cm, clip, width=1.\textwidth]{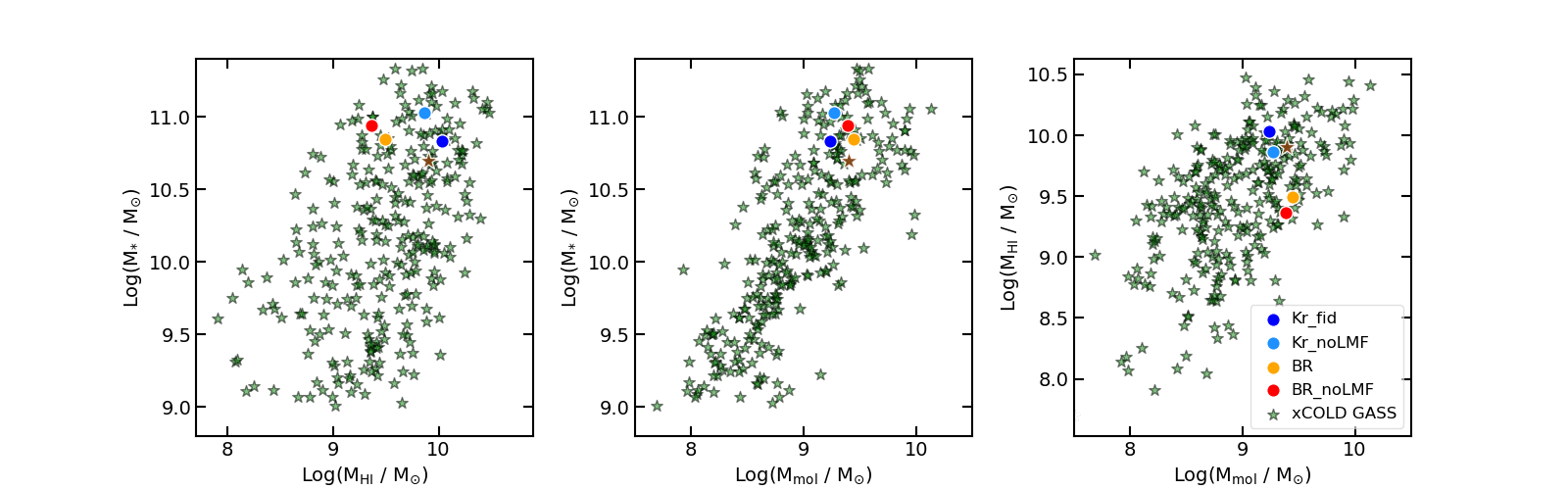} 
\end{minipage} 
\vspace{.5ex}
\caption{Predictions for the stellar-to-HI mass relation ({\sl left panel}), 
stellar-to-molecular mass relation ({\sl middle panel}),  
and HI-to-molecular mass relation ({\sl right panel}) 
from simulations 
\textcolor{Blue}{\bf Kr\_~fid},  \textcolor{Cerulean}{\bf Kr\_noLMF}, 
\textcolor{YellowOrange}{\bf BR}, and \textcolor{Red}{\bf BR\_noLMF}. 
Observations (shown with green stars) represent a sample of SDSS-selected galaxies in the xCOLD GASS survey \citep{Saintonge2017}. 
As a reference, the brown-star symbol shows the location of the MW \citep[according to][]{Kalberla2009, Bland-Hawthorn2016}.}
\label{xCold} 
\end{figure*}

Figure~\ref{jcirc} shows the distribution of the galaxy stellar mass as a function of the circularity of the stellar orbits. The latter quantity is defined via the ratio $J_{\rm z}/J_{\rm circ}$,  namely the specific 
angular momentum in the direction perpendicular to the disc, over that of a reference circular orbit at the considered distance from the galaxy centre \citep{Scannapieco2009}. The mass distribution of each galaxy model is normalised to the galaxy stellar mass and divided by the width of the $J_{\rm z}/J_{\rm circ}$ bin. This kinematic decomposition splits the simulated galaxies into their bulge and disc components: the former is mainly represented by star particles in the peak where $J_{\rm z}/J_{\rm circ}=0$, the latter by those having $J_{\rm z}/J_{\rm circ} \sim 1$.  

Models \textcolor{Blue}{\bf Kr\_~fid} and \textcolor{YellowOrange}{\bf BR} have a reduced bulge component (with respect to \textcolor{Cerulean}{\bf Kr\_noLMF} and \textcolor{Red}{\bf BR\_noLMF},  respectively), LMF suppressing SF at high-z, when the bulge mainly assembles its stellar mass.
The galaxy \textcolor{YellowOrange}{\bf BR} has a more pronounced disc, as a consequence of its high, low-redshift SFR. 

In Figure~\ref{BsuT} we explore the evolution of the galaxy stellar mass as a function of the bulge-over-total stellar mass ratio B/T.  When computing B/T, we assume that counter-rotating (i.e. $J_{\rm z}/J_{\rm circ} < 0$) stellar particles contribute to half of the bulge mass\footnote{Our estimate for the B/T ratio can also include the contribution from bar, satellites, and stellar streams within r$_{\rm gal}$. As a consequence, the B/T values that we quote should not be directly compared with observational photometric estimates. The latter ones are lower than the corresponding kinematic determinations \citep{Scannapieco2010}.}. 
Figure~\ref{BsuT} demonstrates that forming galaxies assemble their stellar mass mainly in their disc below $z \sim 3 - 2$, when the B/T ratio monothonically decreases with redshift. All the models \textcolor{Blue}{\bf Kr\_~fid},  \textcolor{Cerulean}{\bf Kr\_noLMF}, \textcolor{YellowOrange}{\bf BR}, and \textcolor{Red}{\bf BR\_noLMF} become disc-dominated (B/T~$\lesssim 0.5$) below $z \sim 0.2$. 
Stellar masses of the four galaxies at $z=0$ are listed in Table~\ref{tab:GasMasses}, along with the kinematical estimate of their B/T.

\subsection{The star-forming ISM} 
\label{subsec:ISM}

The physical properties of the multi-phase, star-forming ISM determine the formation path of a galaxy.  
To investigate how different prescriptions for the availability of H$_{2}$ shape the final properties of a galaxy, we study the distribution of the actual fuel of SF. Figure~\ref{MmolMaps} shows the distribution of the molecular gas in models \textcolor{Blue}{\bf Kr\_~fid} and \textcolor{YellowOrange}{\bf BR}, at $z=0$. We consider only multi-phase particles having molecular gas.  
The galaxy \textcolor{Blue}{\bf Kr\_~fid} is characterised by a complex ISM: molecular gas is mainly confined within narrow spiral arms and defines a clear spiral pattern, with an intricated network of lanes in the innermost regions. The bulk of molecular gas distributes across a thin disc, whose height never exceeds our force resolution\footnote{Force resolution 
	of our simulations is: $2.7 \varepsilon_{\rm Pl} \sim 1.2$~kpc.  
	The height of the molecular gas disc actually does not exceed 
	$\sim 0.5$~kpc above nor below the galactic plane.}. How features of the \textcolor{Blue}{\bf Kr\_~fid} galaxy resemble observations of local spirals \citep[e.g. those in the PHANGS-ALMA survey,][]{Leroy2021} is striking. 

By contrast, the \textcolor{YellowOrange}{\bf BR} galaxy has a gas disc where the distribution of molecular gas is nearly uniform. This gas homogeneity breaks in two regions: $i)$ where gas accretion is not continuously replenishing the forming disc (at $\sim 10$~kpc from the galaxy centre) and stellar feedback displaces molecular gas, producing gas depressions/holes; and $ii)$ in the outermost regions of the galaxy disc, where spiral arms become thicker as they are sites of ongoing SF \citep[see][for a discussion on the inside-out formation of simulated galaxies adopting the BR model]{Valentini2019}.

Relevant masses of cold and molecular gas for the four models in the reference suite of simulations are listed in Table~\ref{tab:GasMasses}.  As for galaxies \textcolor{Blue}{\bf Kr\_~fid} and \textcolor{YellowOrange}{\bf BR}, we see that \textcolor{Blue}{\bf Kr\_~fid} has a~$\sim 2.1$ times larger M$_{\rm cold}$, but a~$\sim 1.6$ times smaller M$_{\rm mol}$ than \textcolor{YellowOrange}{\bf BR}. 
The \textcolor{Blue}{\bf Kr\_~fid} model predicts a galaxy-wide ratio M$_{\rm mol}$/M$_{\rm cold}$ (i.e. considering gas masses within r$_{\rm gal}$) smaller by a factor of~$\sim 3.5$ with reference to BR.  
Cold gas in \textcolor{Blue}{\bf Kr\_~fid} is converted into molecular with a lower efficiency, and the dependency of $f_{\rm H_{2}}$ on cold gas density and metallicity (see equation~(\ref{eq:f_mol_kr})) produces the complex morphology of the gaseous disc. On the other hand, the model \textcolor{YellowOrange}{\bf BR} shows that cold gas is easily and almost entirely converted into molecular as soon as it gets pressurised,  according to equation~(\ref{eq:f_mol}).  

A quantitative analysis of the evolution of $f_{\rm H_{2}}$ according to the two different prescriptions that we adopt is shown in Figure~\ref{fmolMaps}.  We study how $f_{\rm H_{2}}$ varies as a function of redshift, gas metallicity, and cold gas number density.  In the three panels of Figure~\ref{fmolMaps} we analyse statistical properties of a randomly-selected sub-sample of star-forming particles across cosmic time. To consistently compare predictions from the Kr and BR prescriptions, we contrast: $i)$ $f_{\rm H_{2}}$ for star-forming particles in the \textcolor{Blue}{\bf Kr\_~fid} simulation; and $ii)$ $f_{\rm H_{2}}$ that equation~(\ref{eq:f_mol}) would have produced for the same particles if the BR model had been instead adopted. In this way, we can compare the predictions from the Kr and BR models for star-forming particles with the same physical conditions.  

Depending on the adopted prescription, $f_{\rm H_{2}}$ has either an increasing or decreasing trend as $z$ decreases: the Kr model predicts that $f_{\rm H_{2}}$ almost saturates to $1$ at high z, while it is as low as~$\sim 10^{-3}$ at $z=0$, thus explaining the smaller M$_{\rm mol}$ of \textcolor{Blue}{\bf Kr\_~fid}. Moreover, $f_{\rm H_{2}}$ mildly increases in the \textcolor{Blue}{\bf Kr\_~fid} model as a function of both the gas metallicity and cold gas number density. In particular, we can appreciate that the BR prescription predicts $f_{\rm H_{2}}$ values which are consistent or higher\footnote{Throughout this work, we keep constant the value of $P_0$ 
	within the \citetalias{BR2006} model (see equation~\ref{eq:f_mol}). 
	Increasing $P_0$ can decrease the value of $f_{\rm H_{2}}$. 
	However, we decided here the rely on the fiducial value of $P_0$ calibrated in previous 
	works \citep{Valentini2020} and to focus on the dependence of $f_{\rm H_{2}}$ from 
	the physical properties of the ISM.} than those in the Kr model, depending on the considered cold gas number density. 

Figure~\ref{xCold} demonstrates how final gas masses are sensitive to the adopted SF model and to LMF, and how they relate to the galaxy stellar mass.  We compare predictions from models \textcolor{Blue}{\bf Kr\_~fid},  
\textcolor{Cerulean}{\bf Kr\_noLMF}, 
\textcolor{YellowOrange}{\bf BR}, and \textcolor{Red}{\bf BR\_noLMF} to observations targeting a sample of local SDSS galaxies in the xCOLD GASS survey \citep{Saintonge2017}.  
The figure, specifically, shows: $(i)$ the stellar mass of simulated galaxies within r$_{\rm gal}$; $(ii)$ the HI gas in simulations, estimated as the hydrogen mass in the cold phase minus M$_{\rm mol}${\footnote{We have no direct access to the neutral hydrogen in our simulations. We have assumed that all the hydrogen in the cold phase is neutral to retrieve M$_{\rm cold}$, and that the hot and the cold phases of each multiphase particle share the same chemical composition. We note that if we estimated the neutral gas mass as M$_{\rm HI} = 0.76 \, $M$_{\rm cold}$ - M$_{\rm mol}$ ($0.76$ being the assumed fraction of neutral hydrogen), instead of taking into account each particle cold gas fraction and its actual chemical composition, final quantities would differ from those quoted in Table~\ref{tab:GasMasses} by less than $10$\%.}}; $(iii)$ the molecular gas mass of simulated galaxies, that we assume can be fairly compared to the observed H$_{2}$. All relevant numbers are listed in Table~\ref{tab:GasMasses}.  
Model predictions are in line with observations. 
The comparison suggests that \textcolor{Blue}{\bf Kr\_~fid} and \textcolor{Cerulean}{\bf Kr\_noLMF} agree very well with observed galaxies of similar stellar mass, while \textcolor{YellowOrange}{\bf BR} and \textcolor{Red}{\bf BR\_noLMF} tend to have an HI mass somewhat lower than the average for a given stellar and H$_{2}$ mass. 
We would like to emphasize that even though the assumptions made to estimate HI and H$_{2}$ from our simulations can slightly affect the model position in the panels of Figure~\ref{xCold}, we expect that the satisfactory match with observed data still holds. 

To summarise, the more complex morphology and the molecular gas distribution that the \textcolor{Blue}{\bf Kr\_~fid} simulation shows stem from the joint dependence of $f_{\rm H_{2}}$ on redshift, metallicity, and density that the Kr model assumes.

\subsection{Impact of low-metallicity stellar feedback} 
\label{subsec:EF}

\begin{figure}
\newcommand{\captionfonts}{\small}
\begin{minipage}{\linewidth}
\centering
\includegraphics[trim=0.2cm 0.cm 0.2cm 0.2cm, clip, width=1.\textwidth]{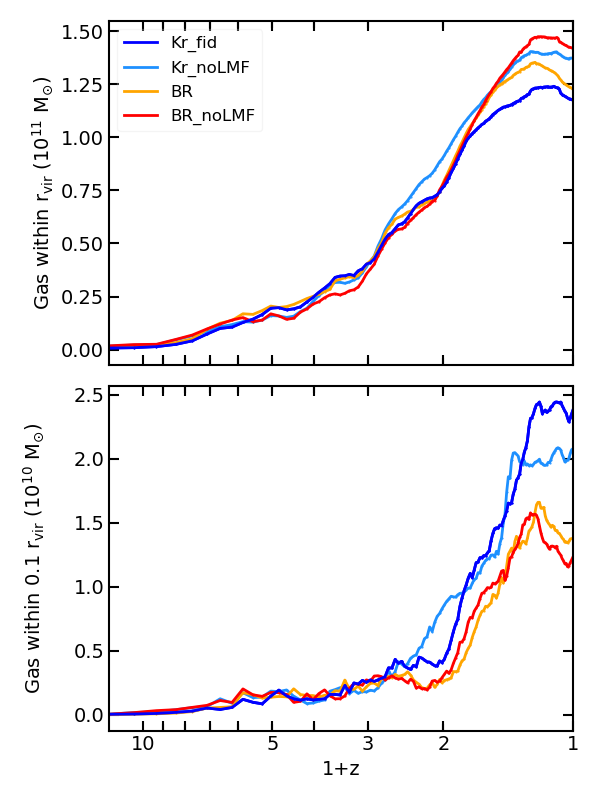} 
\end{minipage} 
\caption{Evolution of the gas mass within~r$_{\rm vir}$ {\sl (top panel)} and within $0.1$~r$_{\rm vir}$ {\sl (bottom panel)} 
for the models \textcolor{Blue}{\bf Kr\_~fid},  \textcolor{Cerulean}{\bf Kr\_noLMF}, 
\textcolor{YellowOrange}{\bf BR}, and \textcolor{Red}{\bf BR\_noLMF}.  
LMF delays gas accretion within~r$_{\rm vir}$, while the conversion of gas into stars at a different rate for {\sl Kr} models contributes to the larger amount of gas within $0.1$~r$_{\rm vir}$ below $z=1$.}
\label{mah} 
\end{figure}

\begin{figure*}
\newcommand{\captionfonts}{\small}
\begin{minipage}{\linewidth}
\centering
\includegraphics[trim=0.cm 4.cm 0.cm 0.cm, clip, width=1.\textwidth]{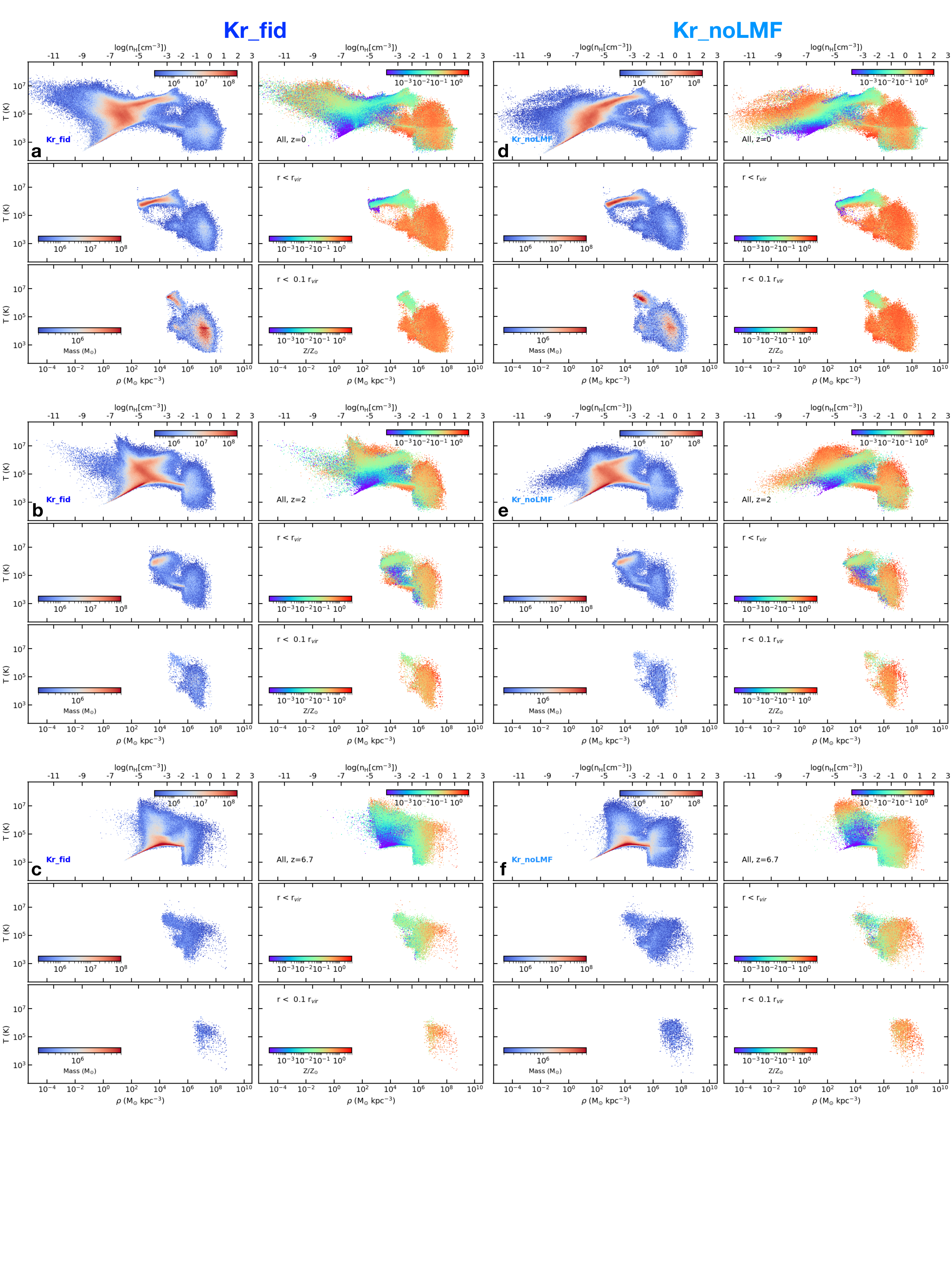} 
\end{minipage} 
\caption{Distribution of gas particles in the density-temperature plane 
	in simulations \textcolor{Blue}{\bf Kr\_~fid} {\sl (panels: a, b, c)} 
	and \textcolor{Cerulean}{\bf Kr\_noLMF} {\sl (panels: d, e, f),} 
	at redshifts $z = 0$ {\sl (top row),} $z = 2$ {\sl (middle)},  and $z = 6.7$ {\sl (bottom)}. 
	The colour encodes either the gas mass per density-temperature bin 
	{\sl (left-hand sub-plots of each panel)} or 
	the mean metallicity per bin {\sl (right-hand ones)}.
	Top sub-plots in each panel show the distribution of all the gas particles 
	in the Lagrangian region,
	middle and bottom sub-plots refer to gas particles within the virial radius r$_{\rm vir}$ and 
	within $0.1$~r$_{\rm vir}$, respectively. 
	While the metallicity color bars in all the panels share minimum and maximum values,  
	the gas-mass limits change with different distances from the galaxy centre, to better capture features.}
\label{PDs} 
\end{figure*}

\begin{figure*}
\newcommand{\captionfonts}{\small}
\begin{minipage}{\linewidth}
\centering
\includegraphics[trim=3.cm 0.cm 3.cm 0.cm, clip, width=1.\textwidth]{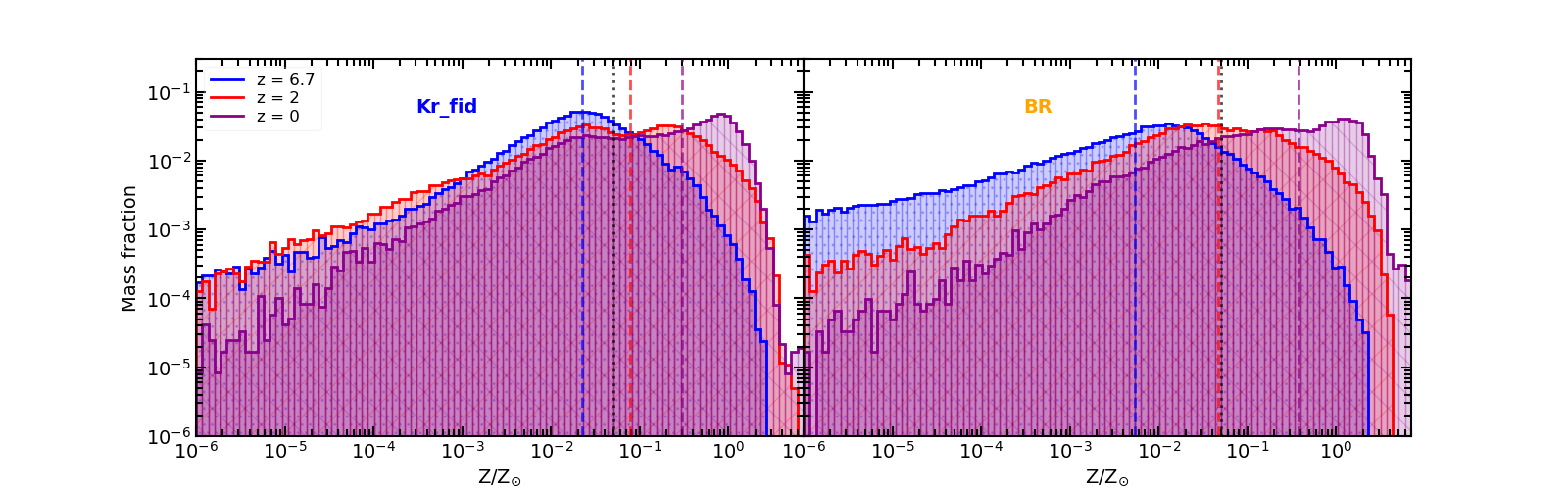} 
\end{minipage} 
\caption{Metallicity distribution functions of all the multi-phase gas particles for models 
	\textcolor{Blue}{\bf Kr\_~fid} {\sl (left-hand panel)} 
	and \textcolor{YellowOrange}{\bf BR} {\sl (right panel)} at different redshift. 
	Dashed blue, red, and purple vertical lines describe the median of the three distributions for 
	each simulation. Median values for \textcolor{Blue}{\bf Kr\_~fid} are: 
	$0.026$, $0.038$, and $0.215$, for $z= 6.7$,~$2$, and~$0$, respectively. 
	As for \textcolor{YellowOrange}{\bf BR}, median values are as follows: 
	$0.005$, $0.043$, and $0.301$. 
	The dotted black vertical line shows the metallicity threshold $Z/Z_{\odot} = 0.05$ for LMF.}
\label{MP_Zhist} 
\end{figure*}

\begin{figure*}
\newcommand{\captionfonts}{\small}
\begin{minipage}{\linewidth}
\centering
\includegraphics[trim=3.cm 0.cm 3.cm 0.cm, clip, width=1.\textwidth]{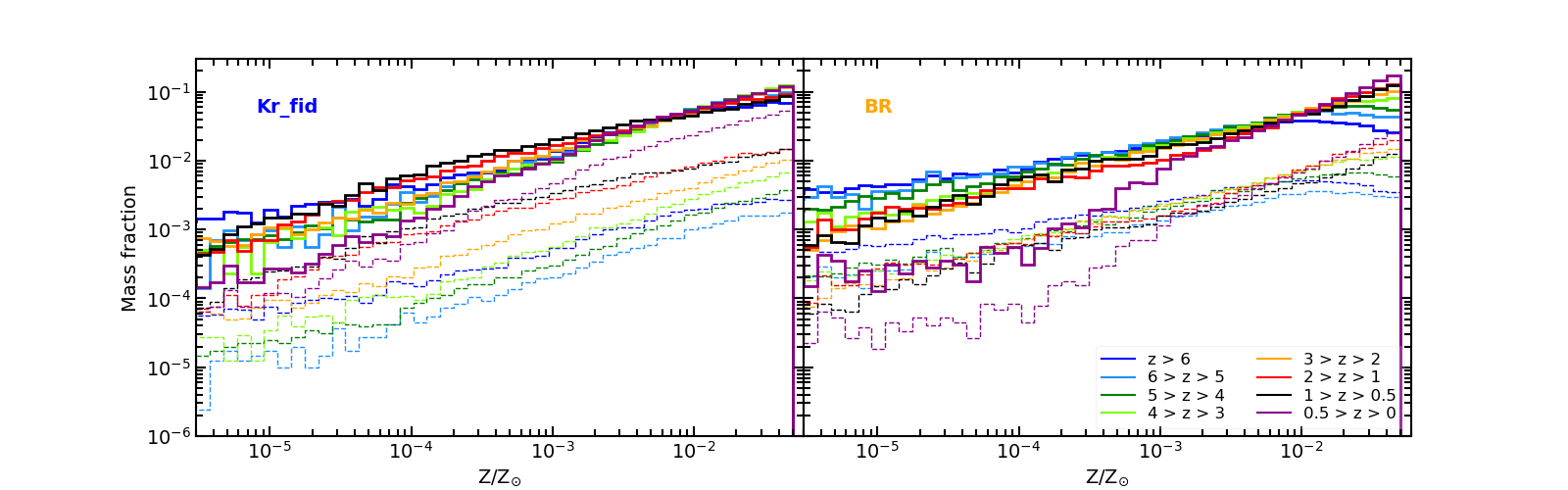} 
\end{minipage} 
\caption{Metallicity distribution functions of all the star-forming gas particles with 
	$Z/Z_{\odot} < 0.05$ contributing to LMF 
	for models \textcolor{Blue}{\bf Kr\_~fid} {\sl (left panel)} 
	and \textcolor{YellowOrange}{\bf BR} {\sl (right panel)}. 
	Different colours refer to different redshifts. 
	Mass fractions are computed over either the gas mass in the considered redshift bin 
	(solid lines) or the total gas mass across all the redshift bins (dashed lines).
	Dashed metallicity distributions are useful to weigh the relative contribution of 
	low-metallicity, star-forming gas and its ensuing LMF across time.}
\label{EF_Zhist} 
\end{figure*}

Figure~\ref{mah} describes the evolution of the mass of gas within r$_{\rm vir}$ and r$_{\rm gal}$ for the four galaxies in the reference set.  Galaxies share comparable gas accretion histories. LMF hampers and delays the accretion of gas from the large-scale environment within r$_{\rm vir}$ for models \textcolor{Blue}{\bf Kr\_~fid} and \textcolor{YellowOrange}{\bf BR} with respect to \textcolor{Cerulean}{\bf Kr\_noLMF} and \textcolor{Red}{\bf BR\_noLMF}, respectively. 
The final mass of gas available results from the galaxy gas accretion, internal gas dynamics (i.e. galactic outflows powered by AGN and SN feedback), gas cooling and SF.  The lower SFR of the \textcolor{Blue}{\bf Kr\_~fid} and \textcolor{Cerulean}{\bf Kr\_noLMF} models at low z contributes to the larger gas mass within their r$_{\rm gal}$. 
Gas masses within r$_{\rm vir}$ and r$_{\rm gal}$ at $z=0$ for the four models are reported in Table~\ref{tab:GasMasses}.  

LMF affects the distribution of gas density and temperature at different redshifts. 
Figure~\ref{PDs} shows the mass and metallicity distribution of gas in the density-temperature phase diagram for models \textcolor{Blue}{\bf Kr\_~fid} and \textcolor{Cerulean}{\bf Kr\_noLMF}, at $z = 6.7$,~$2$,~and~$0$. 
In Figure~\ref{PDs}, the gas particle density is the SPH density; 
as for their temperature, we consider the SPH estimate for single-phase particles and 
the mass-weighted average of the temperatures of the hot and cold phases for multi-phase particles. Multi-phase particles occupy the region of phase diagrams where $\text{log(n}_{\rm H} [\text {cm}^{-3}]) > -2$ (corresponding to n$_{\rm H, \, thres}$), and are allowed to cool down via atomic cooling. The spread in density and temperature that they exhibit is a characteristic feature of the advanced modelling of the ISM within the MUPPI sub-resolution model \citep[see][for details]{Valentini2017}. 

One characteristic signature of LMF is on the hottest (T[K]$ \gtrsim 10^6$) 
and most diffuse ($\text{log(n}_{\rm H} [\text {cm}^{-3}]) \lesssim -5$) gas surrounding the galaxies: when LMF is on,  there is a prominent tail of hot and diffuse gas outside of r$_{\rm vir}$ (both at $z=2$ and $z=0$). This gas in \textcolor{Blue}{\bf Kr\_~fid} is prevented from being easily accreted within r$_{\rm vir}$ (see also Figure~\ref{mah}),  at variance with what happens in the model \textcolor{Cerulean}{\bf Kr\_noLMF}. The origin of this behaviour stems from the energy deposited by LMF in and around the forming galaxy at higher redshift (see e.g. the larger amount of gas heated by LMF at $z=6.7$, at T[K]$ \gtrsim 10^5$ and $\text{log(n}_{\rm H} [\text {cm}^{-3}]) \lesssim -3$): episodes of LMF promote a long-lasting modification of gas cooling properties. By analysing the entropy content of gas particles, we find that particles heated up by LMF locate on a higher adiabatic (than in noLMF models): as a consequence, their cooling time is longer and they are prevented from being easily (re-)accreted towards the centre of the forming galaxy. 

Besides the ability to maintain the ambient gas in a more diffuse state, LMF also leaves an imprint on the multi-phase reservoir of gas. The injections of a larger amount of SN energy hampers SF at high z and reduces the metal content of the star-forming ISM (focus on panels $c$ and $f$ of Figure~\ref{PDs}).  
The metallicity distribution of the gas in the phase diagrams allows us to appreciate how LMF also affects the metal content of the gas in and around galaxies, and plays a crucial role in delaying the enrichment of the ISM.

As for the impact of LMF across cosmic time, Figure~\ref{MP_Zhist} shows the metallicity distribution function of the multi-phase gas in the \textcolor{Blue}{\bf Kr\_~fid} and \textcolor{YellowOrange}{\bf BR} galaxies, at different redshifts (the same considered in Figure~\ref{PDs}). All the distributions have an extended low-metallicity tail, mainly originating from the cosmological accretion of almost-pristine gas. The median of the metallicity distribution function is 0.077 for \textcolor{Blue}{\bf Kr\_~fid} and 0.046 for \textcolor{YellowOrange}{\bf BR},  i.e. close to or still below the metallicity threshold $Z/Z_{\odot} = 0.05$ for LMF even at $z=2$. 

Figure~\ref{EF_Zhist} focuses on metallicity properties of those star-forming particles which have actually provided the ISM with LMF energy, across cosmic time. We analyse a randomly-selected sub-sample of star-forming particles releasing LMF feedback, and split them into different redshift bins to investigate possible statistical trends in their evolution. 
For each galaxy model and redshift bin we consider the metallicity distribution of the gas mass fraction. We express it as either $(i)$ the mass of gas in each metallicity bin over the mass of gas in the entire redshift bin, or $(ii)$ the mass of gas in each metallicity bin over the total mass of star-forming gas providing LMF (i.e. in all the z bins; dashed histograms). 

In the \textcolor{Blue}{\bf Kr\_~fid} galaxy model, the metallicity distribution of gas supplying LMF energy mildly steepens as the redshift decreases, without changing significantly across time. The contribution from LMF is extremely important also at low z, LMF involving an amount of gas which is not negligible if compared to the weakly-enriched gas mass at previous times (e.g. purple, dashed distribution). Interestingly, the metallicity distribution function of gas injecting LMF energy substantially steepens as the redshift decreases in the \textcolor{YellowOrange}{\bf BR} model. There is indeed a significant enrichment of the ISM at low z due to the higher SFR in this galaxy (Figure~\ref{SFH}). 

In summary, the additional source of energy provided by LMF is key in regulating SF at high z and in controlling the amount of gas that can replenish the galaxy reservoir and fuel its SF.  Moreover, not only is LMF important at high z, when the gas is intuitively more metal-poor, but its role is significant throughout the evolution of the galaxy.

\subsection{Galaxy profiles and general properties} 
\label{subsec:profiles}

\begin{figure}
\newcommand{\captionfonts}{\small}
\begin{minipage}{\linewidth}
\centering
\includegraphics[trim=0.cm 1.cm 0.8cm 3.cm, clip, width=1.\textwidth]{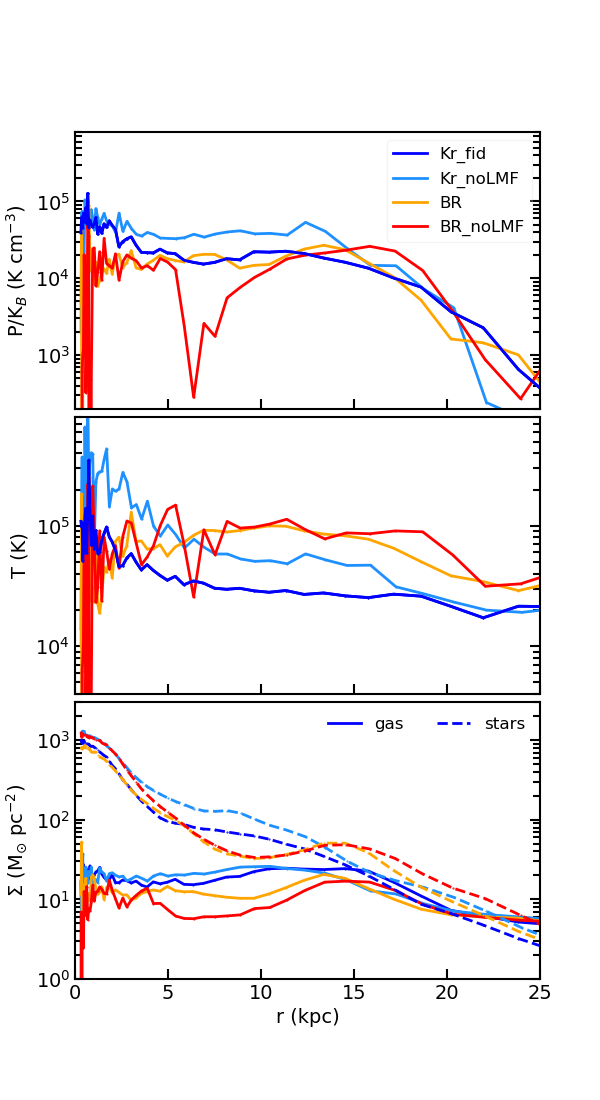} 
\end{minipage} 
\caption{Gas pressure ({\sl top panel}), 
temperature ({\sl middle panel}),  
and gas and stellar surface density ({\sl bottom panel}) radial profiles 
for models \textcolor{Blue}{\bf Kr\_~fid},  \textcolor{Cerulean}{\bf Kr\_noLMF}, 
\textcolor{YellowOrange}{\bf BR}, and \textcolor{Red}{\bf BR\_noLMF}. }
\label{Profiles} 
\end{figure}

\begin{figure}
\newcommand{\captionfonts}{\small}
\begin{minipage}{\linewidth}
\centering
\includegraphics[trim=0.cm 0.cm 0.cm 0.cm, clip, width=1.\textwidth]{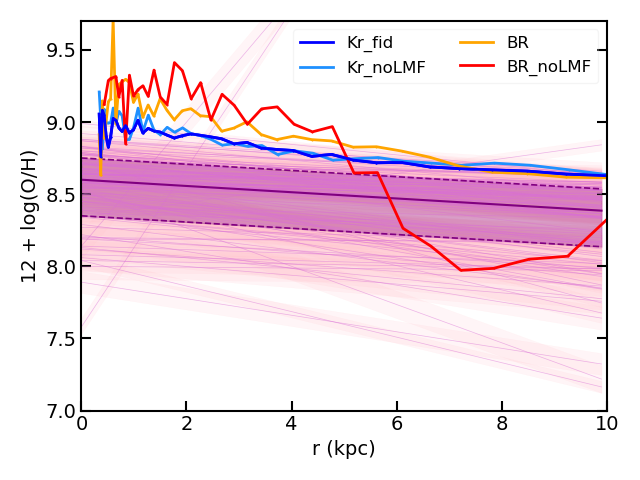} 
\end{minipage} 
\caption{Oxygen abundance radial profiles of gas in the simulated galaxies 
	\textcolor{Blue}{\bf Kr\_~fid},  \textcolor{Cerulean}{\bf Kr\_noLMF}, 
	\textcolor{YellowOrange}{\bf BR}, and \textcolor{Red}{\bf BR\_noLMF}. 
	The purple line shows the median profile from the sample of $130$~nearby, late-type galaxies 
	of \citet{Pilyugin2014}, while the shaded area around it 
	depicts the region between the 16th and 84th percentile (dashed lines). For completeness, profiles of each galaxy in the observational sample are shown in pink, with shaded envelopes showing the scatter of the oxygen abundance around each trend.
}
\label{MetProfiles} 
\end{figure}

Including LMF and assuming different models to estimate H$_{\rm 2}$ contribute to determine final galaxy properties. 
Figure~\ref{Profiles} shows gas pressure, temperature, gas and stellar surface density radial profiles for the reference set of simulated galaxies. 
The discs of galaxies \textcolor{Blue}{\bf Kr\_~fid} and \textcolor{Cerulean}{\bf Kr\_noLMF} are more pressurised than the others, especially in the innermost regions due to the larger amount of gas accreted within r$_{\rm gal}$ (Figure~\ref{mah}).  
Gaseous discs in Kr models are also usually colder (especially at $5 < \text{ r[kpc]} < 15$) than the BR ones. 

The normalization of gas surface density profiles (Figure~\ref{Profiles}, bottom panel) provides additional evidence for the larger amount of gas which is not converted into stars in the Kr models. 
As for the normalization of stellar surface density profiles, \textcolor{Blue}{\bf Kr\_~fid} and \textcolor{YellowOrange}{\bf BR} have slightly-lower central values, since LMF reduces the prominence of the galaxy bulge. 
Fitting the stellar surface density profiles (between $0 < \text{r[kpc]}< 25$) with an exponential law $\Sigma \text{(r)} \propto \text{exp(- r/r$_{\rm s}$)}$, r$_{\rm s}$ being the scale or effective radius, we can find the effective radius of the four galaxies in the reference suite of simulations. They range between $ 4.1 < \text{ r$_{\rm s}$[kpc]} < 4.4 $ and are listed in Table~\ref{tab:GasMasses}.  As a reference, r$_{\rm s} = 3.29 \pm 0.56$~kpc for the thick disc of the MW \citep{mcMillan2011}. 

Figure~\ref{MetProfiles} shows the oxygen abundance profiles of gas in the simulated 
galaxies, at $z=0$ and in observations \citep[][]{Pilyugin2014}. To ensure a fair comparison, we compute the metallicity as 
$12 \, +$ log$_{10} \bigl({\text O}/{\text H} \bigr)$, where 
O and H are the number densities of oxygen (O) and of H, respectively.  
The slopes of the profiles of all the simulated galaxies (but \textcolor{Red}{\bf BR\_noLMF}) are in good agreement with observations.  This comparison highlights the ability of our hydrodynamical scheme and sub-resolution model to properly describe the circulation of metals and the ISM enrichment at different galaxy radii, while suggesting a slight excess of SF and ensuing metal production especially in the inner regions of simulated galaxies. The dip between $5 < \text{ r[kpc]} < 10$ in the \textcolor{Red}{\bf BR\_noLMF} model stems from the gas depression due to the lack of gas replenishment onto the forming galaxy disc (as discussed in Section~\ref{subsec:ISM} and shown in Figure~\ref{StellarGasMaps}).
As for the normalization, profiles of the \textcolor{YellowOrange}{\bf BR} and \textcolor{Red}{\bf BR\_noLMF} models overshoot (by $\gtrsim 0.2$~dex) observations in the innermost regions ($r \lesssim 5$~kpc). They lie above the profiles of \textcolor{Blue}{\bf Kr\_~fid} and \textcolor{Cerulean}{\bf Kr\_noLMF}, as a result of the higher SFR of BR models at low-z. 
How the gas metal content of the ISM of simulated galaxies compares to observations provides further evidence in support of the adoption of the Kr model.

\begin{figure}
\newcommand{\captionfonts}{\small}
\begin{minipage}{\linewidth}
\centering
\includegraphics[trim=0.cm 0.cm 0.cm 0.cm, clip, width=1.\textwidth]{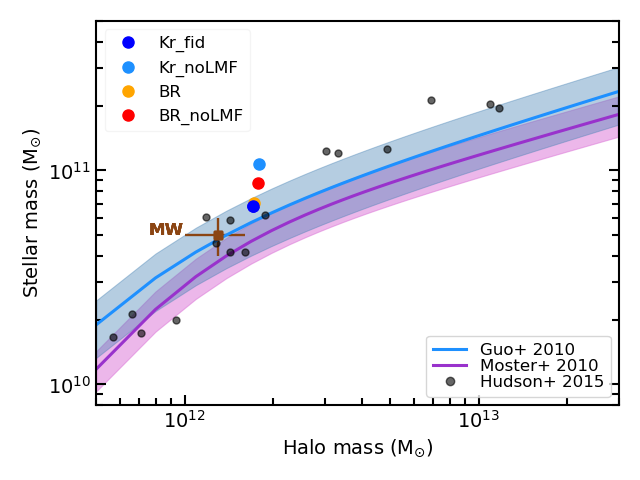} 
\end{minipage} 
\caption{Stellar-to-halo mass relation for galaxies \textcolor{Blue}{\bf Kr\_~fid},  
	\textcolor{Cerulean}{\bf Kr\_noLMF}, \textcolor{YellowOrange}{\bf BR}, 
	and \textcolor{Red}{\bf BR\_noLMF}.
	Gray filled circles show observed galaxies from the CFHT Legacy Survey \citep[][]{Hudson2015}. 
	As a reference, we also show the position of the MW according to \citet{Bland-Hawthorn2016}. 
	The purple solid line shows the relation by \citet{Moster2010}, 
	with 1-$\sigma$ uncertainty (shaded envelope) on the normalization. 
	The light-blue solid curve describes the fit derived by \citet{Guo2010}, 
	along with an interval of $0.2$ dex around it (shaded).}
\label{bce} 
\end{figure}

\begin{figure}
\newcommand{\captionfonts}{\small}
\begin{minipage}{\linewidth}
\centering
\includegraphics[trim=0.cm 0.cm 0.cm 0.cm, clip, width=1.\textwidth]{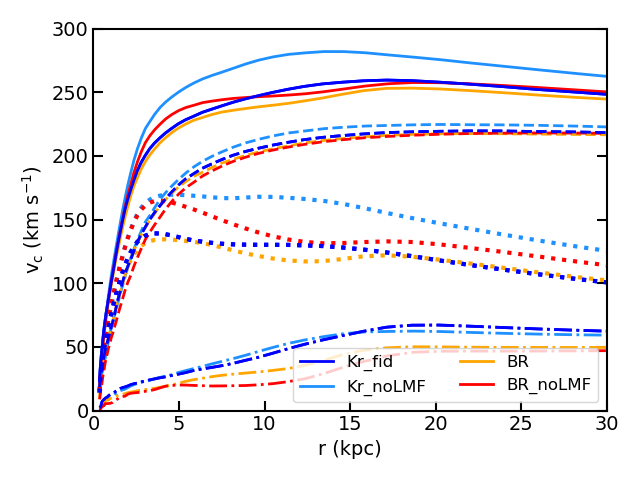} 
\end{minipage} 
\caption{Rotation curves for simulations  \textcolor{Blue}{\bf Kr\_~fid},  
	\textcolor{Cerulean}{\bf Kr\_noLMF}, \textcolor{YellowOrange}{\bf BR}, 
	and \textcolor{Red}{\bf BR\_noLMF}.  
	Solid curves show the circular velocity due to the total mass inside a given radius. 
	Dashed, dotted, and dot-dashed curves describe the contributions from 
	DM, stars, and gas, respectively }
\label{vrot} 
\end{figure}

\begin{figure}
\newcommand{\captionfonts}{\small}
\begin{minipage}{\linewidth}
\centering
\includegraphics[trim=0.cm 0.cm 0.cm 0.cm, clip, width=1.\textwidth]{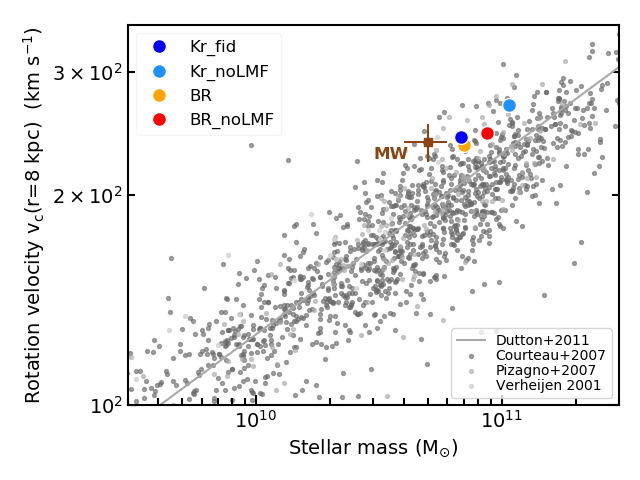} 
\end{minipage} 
\caption{Tully-Fisher relation for models \textcolor{Blue}{\bf Kr\_~fid},  
	\textcolor{Cerulean}{\bf Kr\_noLMF}, \textcolor{YellowOrange}{\bf BR}, 
	and \textcolor{Red}{\bf BR\_noLMF}. 
	Gray filled circles are observational data from samples of spirals by 
	\citet{cou2007, ver2001,piz2007},  
	along with the best fist from \citet{Dutton2011}.
	As a reference, we also show the position of the MW according to \citet{Bland-Hawthorn2016}.}
\label{tf} 
\end{figure}

Figure~\ref{bce} shows the stellar-to-halo mass relation for the four galaxies.  Galaxy stellar masses are listed in Table~\ref{tab:GasMasses}. LMF plays a crucial role in decreasing the galaxy stellar mass, thus improving the comparison with observations \citep{Hudson2015, Bland-Hawthorn2016} and predictions from analytical models \citep{Guo2010, Moster2010}. The final stellar mass of the  the \textcolor{Blue}{\bf Kr\_~fid} galaxy model is the closest to the MW stellar mass \citep{Bland-Hawthorn2016}, that we also show as a reference. 

Figure~\ref{vrot} analyses the rotation curve of the simulated galaxies. We split the contributions from DM, gas, and stellar components to the total circular velocity (v$_{\rm c}$(r) = (G M(<r) /r )$^{1/2}$).  
All the rotation curves are characterised by a steep increase in the innermost regions where the stellar component dominates, while the profiles flatten and decline at large radii, where the contribution from DM becomes dominant.  Differences among the gas and stellar contributions of the various galaxy models reflect different density radial profiles (Figure~\ref{Profiles}, bottom panel). By comparing the rotation curves of our simulated galaxies to that of the MW \citep[e.g.][]{Li2022},  we see that the total rotation curve of the MW exhibits an almost flat behaviour starting from smaller radii ($r \gtrsim 4$~kpc), while the radius at which baryons and DM equally contribute to the total circular velocity is larger than in our models. Indeed,  at the Solar radius $r_{\odot} = 8.15$~kpc \citep{Reid2019}, the total circular velocity of the MW is $\sim 235$~km~s$^{-1}$ \citep{Sofue2009, Li2022}. At that distance, v$_{\rm c, bar}$(r$_{\odot}$)~$\sim 170$~km~s$^{-1}$ (M$_{\rm MW, bar}$(<r$_{\odot}$)~$\sim 4.9 \times 10^{10}$~M$_{\odot}$) and v$_{\rm c, DM}$(r$_{\odot}$)~$\sim 160$~km~s$^{-1}$ (M$_{\rm MW, DM}$(<r$_{\odot}$)~$\sim 4.7 \times 10^{10}$~M$_{\odot}$) \citep{Portail2017, Cautun2020}.  Our simulated galaxies are dominated by DM starting from smaller distances from the centre (with respect to the MW): this might be ascribed to a different past history of the feedback which did not expand the halo as effectively as in the MW.  

Figure~\ref{tf} shows the Tully-Fisher relation \citep{TullyFisher}, cast in terms of rotation velocity versus galaxy stellar mass. To be consistent with observational estimates,we measure the circular velocity at a distance of $8$~kpc from the galaxy centre. We verified that the actual distance at which the circular velocity is evaluated does not play a significant role. Predictions from simulations lie on the upper edge of the region of the plane outlined by observations \citep{cou2007, ver2001,piz2007} and hint at models which are slightly more centrally concentrated than observations. We note that also the MW lies slightly offset with respect to the bulk of the observational sample. 
All the extensive tests that we have carried out to tune the parameters of our sub-resolution model suggest that it is difficult to shift simulated galaxies simultaneously towards smaller stellar masses and lower rotation velocities in the Tully-Fisher plane.  

As a final note, we mention that all the simulations that we have discussed so far include the presence of BHs and AGN feedback. Final properties of BHs are not significantly affected by LMF nor by the different prescriptions assumed to model H$_{\rm 2}$, and they remain in keeping with those of the fiducial model introduced in \citet{Valentini2020}. The impact of AGN feedback on final results is discussed in Appendix~\ref{AppB}. 

Moreover, as for the impact of the numerics on the robustness of the results discussed so far, we find that differences among runs driven by different physical processes (i.e. SF prescriptions and inclusion of LMF) are not comparable to those possibly introduced by the chaoticity of the system. To estimate the latter, we repeated the same run several times, introducing tiny perturbations in our ICs (see Appendix~\ref{AppD} for details).

\section{Discussion} 
\label{sec:discussion}

\subsection{Estimating the molecular gas fraction} 
\label{discussion_H2}

Quantifying the amount of H$_{\rm 2}$ is key to assess the mass of the reservoir of gas which fuels SF. From a numerical perspective, a few approaches have been pursued so far. 

The H$_{\rm 2}$ can be computed by solving the full set of time-dependent, out-of-equilibrium rate equations \citep[e.g.][]{Maio2007}. Although this approach represents the most accurate way to calculate H$_{\rm 2}$, this path is computationally demanding. As a consequence, cosmological simulations relying on these computations are usually limited to small volumes or even to single galaxies, and/or to high-z \citep[e.g.][]{Schaebe2020}.
The formation and dissociation of H$_{\rm 2}$ via non-equilibrium chemistry can be consistently accounted for by packages like KROME \citep{Grassi2014} or GRACKLE \citep{Smith2017}, which can be embedded in simulation codes \citep[e.g.][]{Lupi2018}.

Another possibility is to compute f$_{\rm H_{\rm 2}}$ from quantities that are locally accessible in simulations, e.g. gas density and metallicity \citep{Krumholz2008, Kr2009, Krumholz2009ApJ693, McKee2010, KrumholzGnedin2011}. The aforementioned models feature a spherical molecular complex embedded in a medium with ionizing, Lyman-Werner photons: the f$_{\rm H_{\rm 2}}$ and the radiation field can be thus retrieved from the gas column density and metallicity, once the ISM is made up of a cold and a warm phase in equilibrium. 
Interestingly, \citet{KrumholzGnedin2011} compared the performance of an analytic expression to retrieve f$_{\rm H_{\rm 2}}$ among the aforementioned ones and a highly-accurate, time-dependent chemistry and radiative transfer model when embedded in a cosmological simulation. The latter model included the formation of H$_{\rm 2}$ both on dust grains and in a primordial environment, the abundances of the relevant atomic and molecular species being tracked self-consistently on-the-fly. They demonstrated that the two formulations agree well for metallicity Z~$\gtrsim 0.1$~Z$_{\odot}$.
By adopting one among the different flavours of the aforementioned prescriptions for f$_{\rm H_{\rm 2}}$, \citet{Kuhlen2012} showed how H$_{\rm 2}$-based prescriptions for SF control the SF efficiency directly: this is at variance with the indirect regulation of the cold gas reservoir with stellar feedback. 

\citet{Gnedin2009} introduced a phenomenological model which accounts for the non-equilibrium formation of H$_{\rm 2}$ on dust grains, also including an effective treatment of its self-shielding and the shielding by dust from the dissociating ionizing radiation. They showed how the transition from neutral H to H$_{\rm 2}$ is highly dependent on gas metallicity, a quantity that in their model also contributes to determine dust. 
In a similar effort to refine prescriptions for SF adopted in cosmological simulations, \citet{Christensen2012} proposed a model where non-equilibrium H$_{\rm 2}$ abundances result from the integration over the H chemical network.  Their model includes the formation of H$_{\rm 2}$ on dust grains, shielding, and photoionization.  They also discuss how computing non-equilibrium f$_{\rm H_{\rm 2}}$ is essential in models that feature H$_{\rm 2}$ formation on dust grains: indeed, since the latter process is very inefficient, assigning an instantaneous f$_{\rm H_{\rm 2}}$ to a resolution element would not be correct. Their simulation results at $z=0$ match the observed Tully-Fisher and Kennicutt-Schmidt relations. 
The works by \citet{Pelupessy2006, GnedinKravtsov2011, Feldmann2011, Tomassetti2015} fit in a similar framework.  Interestingly, \citet{GnedinKravtsov2011} proposed a convenient fitting formula to capture the average dependence of f$_{\rm H_{\rm 2}}$ on the ionizing flux and on the dust-to-gas ratio in simulations.

\citet{muppi2010} exploited the phenomenological, pressure-based prescription by \citetalias{BR2006} to include an H$_{\rm 2}$-driven SF in their simulations, resorting to the sub-resolution model MUPPI. \citet{Monaco2012} showed that simulations based on this sub-grid prescription were able to reproduce the Kennicutt-Schmidt relation, without assuming a Schmidt law to compute the SFR.

Interestingly, \citet{Hopkins2013} quantified the impact of assuming different SF criteria in high-resolution ($\sim 500$~M$_{\odot}$ and $\sim 5$~pc for the mass and softening of the gas particles, respectively) simulations of isolated disc galaxies (a MW-like galaxy at $z=0$ and a starburst disc at high redshift). Among the considered criteria, they compared the effect of allowing SF in dense (above a density threshold) gas or only from molecular gas \citep[relying on][to estimate the molecular gas fraction]{KrumholzGnedin2011}. They found that, while the adopted prescription for SF does not play a crucial role in determining the total SFR of the galaxy, it does affect the spatial distribution of the star-forming gas in each simulation. As also confirmed by \citet{Hopkins2014, Hopkins2017} with the FIRE suite, tying the SFR to the molecular phase does not impact the galaxy total SFR (provided that a high SF density threshold is adopted, i.e. $n_{\rm H, \, thres}  \sim 10^3$~cm$^{-3}$), as SF is feedback-regulated in their simulations. At variance with these results, our investigation suggests that the methodology adopted to estimate the available H$_{\rm 2}$ at the sub-resolution level and hence the SFR can play a significant role.  

This section is focused on numerical simulations on purpose. As for semi-analytical models, a few works have quantified the impact of assuming different prescriptions to estimate H$_{\rm 2}$ on galaxy properties \citep[e.g.][]{Fu2010, Lagos2011, Popping2014, Somerville2015, Xie2017}.
Such works agree on the finding that assuming different prescriptions to compute H$_{\rm 2}$ does not have remarkable effect on the global properties (e.g. stellar mass, SFR, metallicity) of galaxies.  They conclude that, on average, galaxy SFHs are not significantly sensitive to different SF laws. Among the aforementioned numerical works based on semi-analytics, some of them included also the \citetalias{BR2006} and the \citetalias{Kr2009} in their comparison set.
However, despite having a valuable predictive power, semi-analytic models do not fully capture gas dynamics as hydrodynamical simulations do: this is a crucial feature within a context where gas accretion, ejection, and circulation can affect the distribution and availability of the multi-phase gas. Moreover, a key piece of information that semi-analytical models cannot access in detail is galaxy morphology. The latter is a fundamental prediction of cosmological simulations like the ones introduced in this work: it can be compared to observations in the attempt to discriminate among possible numerical simplifications of the SF law.

\subsection{Varying the stellar feedback efficiency} 
\label{discussion_EF}

The term early feedback \citep{Stinson2013} originally refers to a purely thermal feedback operating before SN explosions. This form of feedback is meant to provide the ambient medium with radiation energy from young, massive stars, just before they explode as SNe~II. Early stellar feedback represents a UV ionization source, that increases the surrounding gas temperature and supplies heating and pressure support. This feedback channel indeed pre-processes the star-forming ISM and facilitates the effectiveness of SNe at regulating SF. In their simulations, \citet{Stinson2013} show how early stellar feedback helps in suppressing SF at high z.

The early feedback by \citet{Stinson2013} consists thus in the pre-heating of gas by progenitors of SNe~II. 
Inspired by this work and by how early stellar feedback operates in the NIHAO simulations \citep[e.g.][]{Wang2015, Obreja2019}, we explored whether it is possible to account for the feedback by progenitors of local SNe~II at earlier times. Our LMF is however different from the original early feedback. The goal of our LMF is to capture the effect of very massive and energetic SNe exploding in an almost pristine ISM: we thus aim at mimicking the explosions of hypernovae and SNe~II of population III stars \citep[e.g.][for reviews]{Bromm2004, Ciardi2005}. 

The idea of differentiating the outcome of stellar feedback according to the physical properties of the star-forming ISM has been already pursued in cosmological simulations, and a few effective prescriptions have been proposed.

\citet{Maio2010} first modelled the simultaneous evolution of different stellar populations in cosmological simulations. Their runs feature stellar feedback dependent on the underlying gas metallicity and stellar population, where the injected energy and the assumed stellar yields depend on Population III versus Population II regimes, i.e. pair-instability SNe versus SNe.  Interestingly, \citet{Maio2016} included the effect of self-consistently coupled radiative transfer, concluding that radiative feedback from massive Population III stars could be a viable justification for powerful feedback in pristine environments. 

A metallicity- and density-dependent stellar feedback efficiency has been introduced in the Eagle simulation \citep{Schaye2015, Crain2015}. There, the aforementioned efficiency decreases with gas metallicity while increasing with gas density. The adoption of a similar parametrization has a twofold reason: $i)$ radiative losses are expected to increase with increasing metallicity; $ii)$ energy losses in high-density, star-forming regions can make the stellar feedback too inefficient, and have to be counterbalanced. 
Being the gas metallicity lower at higher redshifts,  when also higher densities are usually reached in the star-forming ISM,  such a prescription yields a redshift-dependent stellar feeback efficiency. The latter ranges between asymptotic values, which are parameters of the model tuned to reproduce low-redshift observables, e.g. the galaxy stellar mass function.  In \citet{Valentini2017}, we demonstrate how the SFH of a MW galaxy is affected by variations in the feedback efficiency when a stellar feedback model inspired by that of the Eagle simulation is adopted.

A similar parametrization is adopted in the IllustrisTNG simulation \citep{Pillepich2018}, too. 
They assume that the wind energy available to a star-forming resolution element is metallicity-dependent.  SNe II release indeed a feedback energy which spans the range $(0.9 - 3.6) \times 10^{51}$~erg, according to the metallicity of the star-forming gas cell in which they are expected to explode -- the lower the metallicity, the larger the energy budget.  \citet{Pillepich2018} demonstrate how the metallicity-dependent stellar feedback energy modulation has an impact both on the stellar-to-halo mass relation and on the galaxy stellar mass function, final results strongly depending on the tuning of the parameters.  They show that the stellar mass of a $\sim 10^{12}$~M$_{ \odot}$ halo at $z=0$ can vary by up to a factor of $\sim 2$.

Our effort fits into this framework. However, we decided not to impose a functional form to the stellar feedback efficiency: this is loosely constrained by physics and mainly driven by the need of having a convenient formula for calibrations in simulations. Rather, we explored the possibility of accounting for the effect of the explosion of hypernovae and SNe II in weakly-enriched environments. Our current modelling is effective and only captures the outcome of a complex phase in the life of early stars. It will be improved and refined in the upcoming future.

\section{Summary and Conclusions}
\label{sec:conclusions}

In this paper, we investigated how the reservoir of molecular gas (H$_{2}$) depends on the physical properties of the star-forming ISM and determines the star formation history (SFH) of the forming galaxy. 
We carried out cosmological hydrodynamical simulations of disc galaxies targeting a MW-sized halo. We used the sub-resolution model MUPPI to describe a multi-phase ISM and the physical processes happening at unresolved scales. 

The simulations introduced in this work have been designed to investigate: 
\begin{itemize}
\item the impact of different prescriptions to compute H$_{2}$ on galaxy formation and evolution. We considered: $i)$ the theoretical model by \citetalias{Kr2009} and $ii)$ the phenomenological prescription by \citetalias{BR2006}; 
\item the effect that low-metallicity stellar feedback (LMF) has on the physical properties of the star-forming ISM. As for our effective modelling of LMF, we assume that SNe II exploding in a low-metallicity ($Z < 0.05$~Z$_{\odot}$) ISM inject a larger ($10x$) amount of stellar feedback energy.
\end{itemize}

The most relevant results of this paper can be summarised as follows.
\begin{itemize}
\item  Connecting the SFR to the galaxy reservoir of H$_{2}$ is crucial to properly investigate galaxy formation from a numerical perspective. 
\item Different numerical prescriptions adopted to estimate the fraction of molecular gas, $f_{\rm H_{2}}$, determine final properties of simulated galaxies. While galaxies can have a clear disc-dominated morphology independently of the details entering the modelling of SF, the prominence of their stellar bulge, the spatial extension of their disc, and features of the spiral pattern are a distinctive signature of the adopted model.  
\item The properties of our simulated galaxies (above all, their morphology) suggest that the \citetalias{Kr2009} should be preferred to the \citetalias{BR2006} prescription to estimate $f_{\rm H_{2}}$. 
\item The molecular gas distribution and the more complex morphology that galaxies adopting the  \citetalias{Kr2009} model show originate from the joint dependence of $f_{\rm H_{2}}$ on metallicity,  density, and thus redshift that this model assumes.  
In particular, \citetalias{Kr2009} predicts an $f_{\rm H_{2}}$ which is larger at high z and lower at low z than \citetalias{BR2006}. 
\item The additional source of energy provided by LMF is key in regulating SF at high z and in controlling the amount of gas that can replenish the galaxy reservoir and fuel its SF.  LMF plays a crucial role not only at high z, when the gas is intuitively more metal-poor, but also throughout the evolution of the galaxy.  
\item LMF is able to reduce the stellar mass of the galaxy bulge, by suppressing SF at high z. Its impact on the stellar mass assembly of bulges in late-type galaxies can be even stronger than the influence of AGN feedback.
\item The conclusion that LMF controls the SFH above $z \gtrsim 1$, while the SF model is responsible for the differences in the SFRs especially at lower redshift is robust against the nuisance produced by numerics and the chaotic behaviour of the code.
\end{itemize}

The numerical modelling of the SF process in cosmological simulations can be further improved, by also profiting from the results of very-high resolution ($\sim$~pc) hydrodynamical simulations of portions of the ISM \citep[e.g.][]{Padoan2012, Girichidis2016, Kim2020}. Another interesting direction of investigation is represented by the possibility to consistently account for the formation of H$_{2}$ on dust grains at the sub-resolution level, in those cosmological simulations that feature the formation and evolution of dust \citep[e.g.][and references therein]{Granato2021}.
In this paper, we have focused on the fuel of SF, i.e. the molecular gas. An accurate modelling of the other components entering the computation of the SFR, i.e. the depletion timescale for SF and the SF efficiency, is key as well.  

As a caveat, we note that calibrating the prescriptions adopted to model SF in cosmological simulations by means of observations of nearby galaxies may represent an issue. Assuming that formulations valid in the local Universe can be extrapolated at high redshift is likely at the origin of the inability of most of cosmological simulations to produce galaxies in proto-clusters whose SFRs cannot reach the values suggested by some observations \citep[as discussed, for instance, in][]{Granato2015, Bassini2020}. This calls for the need to further improve the sub-resolution modelling of SF. Taking advantage from recent and upcoming observations of the high-z Universe would be of paramount importance.


\appendix

\section{Tailoring low-metallicity stellar feedback}
\label{AppA}

\begin{figure*}
\newcommand{\captionfonts}{\small}
\begin{minipage}{\linewidth}
\centering
\includegraphics[trim=0.2cm 0.05cm 0.2cm 0.2cm, clip, width=.52\textwidth]{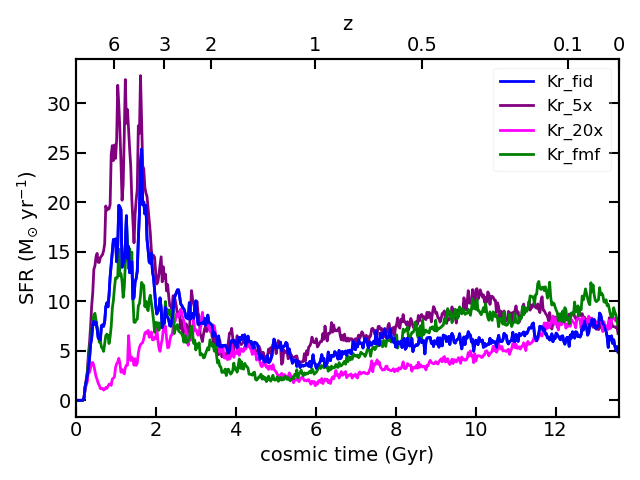} 
\includegraphics[trim=0.2cm 0.cm 0.2cm 0.2cm, clip, width=.47\textwidth]{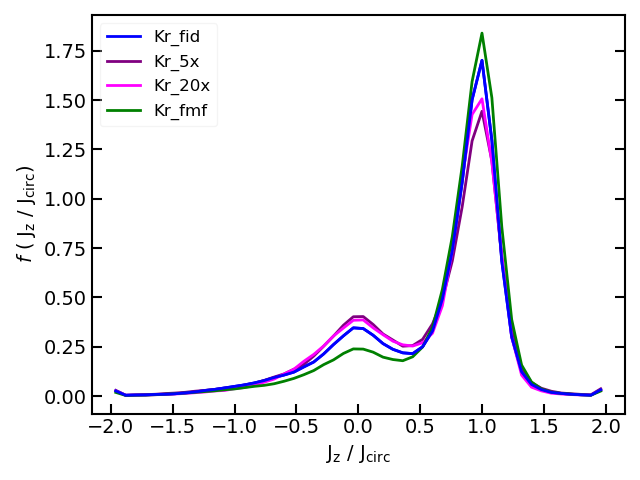}
\end{minipage} 
\caption{Star formation history {\sl (left-hand panel)} 
and kinematic decomposition of the stellar mass within $0.1$~r$_{\rm vir}$ {\sl (right-hand panel)} 
of simulated galaxies 
\textcolor{Blue}{\bf Kr\_~fid}, \textcolor{Plum}{\bf Kr\_5x}, 
\textcolor{WildStrawberry}{\bf Kr\_20x} and \textcolor{LimeGreen}{\bf Kr\_fmf}.}
\label{figgAppA} 
\end{figure*}

This appendix focuses on the tuning of the parameter entering the modelling of LMF, i.e. the boost factor for stellar feedback energy (see Section~\ref{sec:EF}). 
The left panel of Figure~\ref{figgAppA} shows the SFH of models 
\textcolor{Plum}{\bf Kr\_5x}, \textcolor{WildStrawberry}{\bf Kr\_20x}, 
and \textcolor{LimeGreen}{\bf Kr\_fmf} 
(see Table~\ref{tab:sims} and Section~\ref{sec:sims} for details), along with the reference 
\textcolor{Blue}{\bf Kr\_~fid}.  Stellar and gas density maps of the aforementioned models are shown in Figure~\ref{StellarGasMaps}. 

The larger the amount of LMF energy injected, the larger is the delay and suppression of the SF at high z. The clear trend that the models 
\textcolor{Plum}{\bf Kr\_5x}, \textcolor{Blue}{\bf Kr\_~fid}, \textcolor{WildStrawberry}{\bf Kr\_20x} define at high redshift is not evident at low z anymore, due to the integrated effect of the interplay between SF and stellar feedback across time.
Should even less LMF energy than in \textcolor{Plum}{\bf Kr\_5x} be injected, the model \textcolor{Cerulean}{\bf Kr\_noLMF} would be recovered. 
We find that letting low-metallicity star-forming particles supply $10x$ stellar feedback energy (i.e. \textcolor{Blue}{\bf Kr\_~fid}) results in an effective description of LMF: in this way, we can reduce the high-z SF burst, while still promoting the formation of an extended stellar disc. A more accurate modelling of different SF modes associated to the physical properties of star-forming particles is the subject of an upcoming paper. 

In the left panel of Figure~\ref{figgAppA}, the model \textcolor{LimeGreen}{\bf Kr\_fmf} quantifies the impact of assuming a fixed H$_{\rm 2}$ fraction (f$_{\rm H_{\rm 2}} = 0.95$) for low-metallicity star-forming particles. According to this model, not only inject particles with $Z < 0.05$~Z$_{\odot}$ a larger feedback energy, but they also have an f$_{\rm H_{\rm 2}}$ close to unity, thus mimicking a sort of H$_{\rm 2}$ shielding.  Results shown in  Figure~\ref{figgAppA} suggest that there are no strong reasons for the latter simulation to be preferred to \textcolor{Blue}{\bf Kr\_~fid}.

The right panel of Figure~\ref{figgAppA} analyses the distribution of the stellar mass as a function of the circularity of stellar orbits (see the description of Figure~\ref{jcirc} for details) for the same set of simulations.  
Bulge-over-total mass ratios for models \textcolor{Plum}{\bf Kr\_5x}, \textcolor{WildStrawberry}{\bf Kr\_20x}, and \textcolor{LimeGreen}{\bf Kr\_fmf} are B/T$=0.40$, $0.38$, and $0.26$, respectively. 
Interestingly, the model \textcolor{LimeGreen}{\bf Kr\_fmf} exhibits the smallest B/T among all the simulated galaxies and has both the gaseous and stellar discs extended and with a defined spiral pattern (see Figure~\ref{StellarGasMaps}).


\section{Effect of BHs and AGN feedback}
\label{AppB}

\begin{figure*}
\newcommand{\captionfonts}{\small}
\begin{minipage}{\linewidth}
\centering
\includegraphics[trim=0.2cm 0.05cm 0.2cm 0.2cm, clip, width=.52\textwidth]{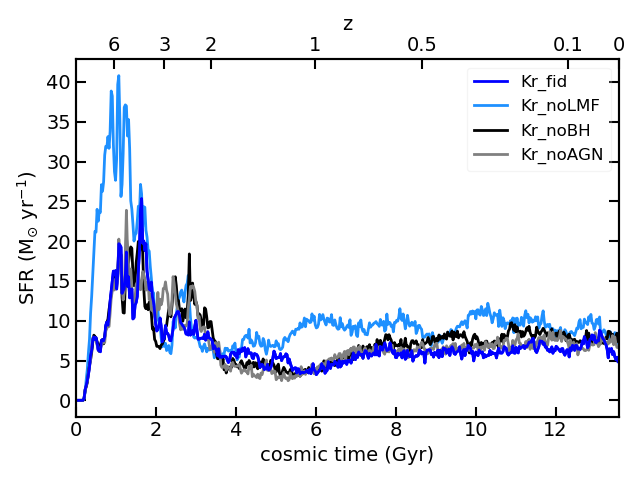} 
\includegraphics[trim=0.2cm 0.cm 0.2cm 0.2cm, clip, width=.47\textwidth]{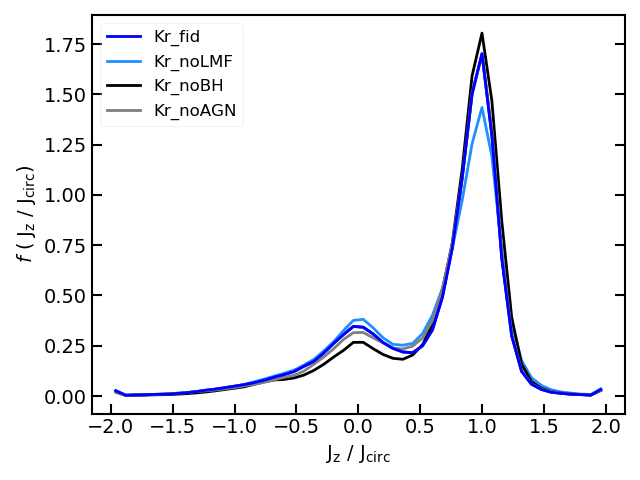}
\end{minipage} 
\caption{Star formation history {\sl (left panel)} and 
kinematic decomposition of the stellar mass within $0.1$~r$_{\rm vir}$ {\sl (right panel)} 
of galaxy models 
\textcolor{Blue}{\bf Kr\_~fid},  \textcolor{Cerulean}{\bf Kr\_noLMF}, 
\textcolor{Black}{\bf Kr\_noBH} and \textcolor{Gray}{\bf Kr\_noAGN}. }
\label{figgAppB} 
\end{figure*}

The simulations considered in this appendix address the questions of $(i)$ whether AGN feedback has an impact on the central compactness of simulated galaxies and $(ii)$ whether assuming that a central, accreting BH were not there can help in reducing the galaxy compactness. 
The reference simulation \textcolor{Blue}{\bf Kr\_~fid} has a central BH of mass $1.67 \times 10^6$~M$_{\odot}$. 
The left panel of Figure~\ref{figgAppB} shows the SFH of models 
\textcolor{Black}{\bf Kr\_noBH} and \textcolor{Gray}{\bf Kr\_noAGN} 
(see Table~\ref{tab:sims} and Section~\ref{sec:sims} for details), along with that of 
\textcolor{Blue}{\bf Kr\_~fid} and \textcolor{Cerulean}{\bf Kr\_noLMF}, as a reference.  
The lower SFR of \textcolor{Black}{\bf Kr\_noBH} with reference to \textcolor{Blue}{\bf Kr\_~fid} between $2 \gtrsim z \gtrsim 1$ can be explained in terms of the AGN-induced over-pressurization of the ISM \citep{Valentini2020}.  The simulations in the left panel of Figure~\ref{figgAppB} -- all adopting the Kr model -- confirm that BH activity can promote episodes of both negative and positive AGN feedback at low redshift in disc galaxies, regardless of the numerical prescription adopted to estimate the H$_{\rm 2}$ 
\citep[see Dicussion in Section~5.3 of][for details]{Valentini2020}.

The right panel of Figure~\ref{figgAppB} analyses the circularity diagram for the same set of simulations.
The stellar mass within r$_{\rm gal}$ at $z=0$ is 
$\text{M}_{\ast} = 7.59 \times 10^{10}$~$\text{M}_{\odot}$ for \textcolor{Black}{\bf Kr\_noBH}, 
while \textcolor{Gray}{\bf Kr\_noAGN} has $\text{M}_{\ast} = 7.10 \times 10^{10}$~$\text{M}_{\odot}$. 
Their bulge-over-total mass ratio is B/T$=0.27$ and B/T$=0.32$, respectively.  

In conclusion, including the central BH and its ensuing AGN feedback reduces the galaxy stellar mass by $15$\%. Assuming that the central, accreting BH is not there does not help in significantly reducing the galaxy compactness nor in shifting the simulated galaxy towards a lower rotation velocity on the Tully-Fisher plane, the model \textcolor{Black}{\bf Kr\_noBH} having v$_{\rm circ}$(r=8~kpc)$=247.43$~km~s$^{-1}$. 
Moreover, we find that LMF is way more crucial than AGN feedback in controlling the stellar mass of the galaxy bulge, in our simulations of disc galaxies.


\section{NIHAO galaxies}
\label{AppC}

\begin{figure*}
\newcommand{\captionfonts}{\small}
\begin{minipage}{\linewidth}
\centering
\vspace{-2.ex}
\includegraphics[trim=0.cm 23.5cm 1.cm 0.cm, clip, width=1.\textwidth]{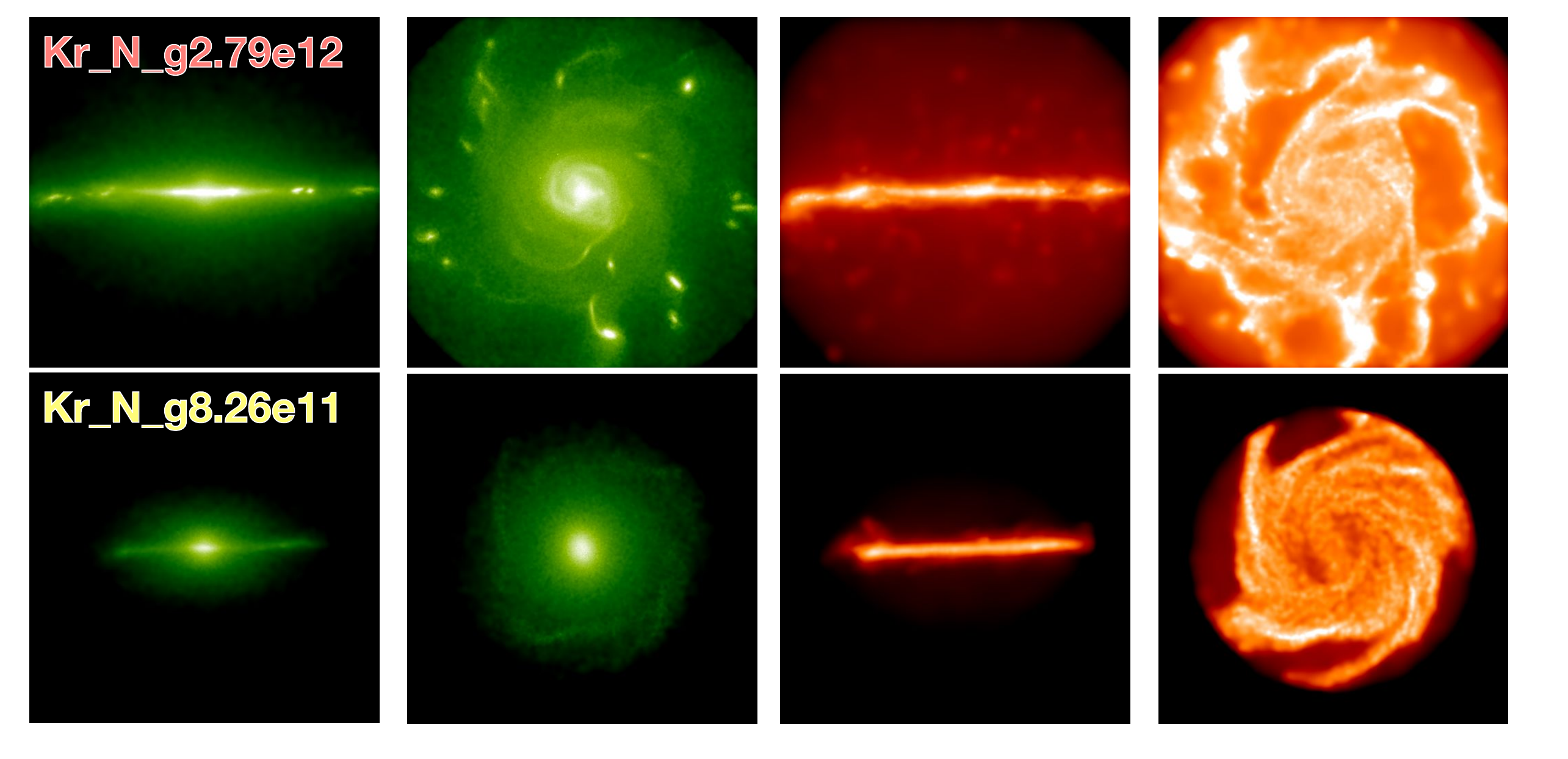} 
\end{minipage} 
\vspace{-2.ex}
\caption{Projected stellar {\sl (first and second columns)} and gas {\sl (third and forth columns)} density maps for the simulated galaxies of the Nihao set, at~$z = 0$. 
First and third columns show edge-on galaxies, second and forth ones depict face-on maps. Each box is $50$~kpc a side.  See Figure~\ref{StellarGasMaps} for comparison. }
\label{StellarGasMaps_Nihao} 
\end{figure*}

\begin{figure*}
\newcommand{\captionfonts}{\small}
\begin{minipage}{\linewidth}
\centering
\includegraphics[trim=0.2cm 0.05cm 0.2cm 0.2cm, clip, width=.52\textwidth]{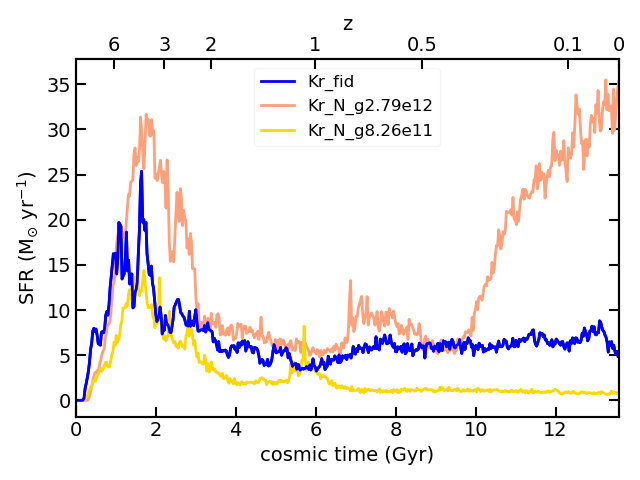} 
\includegraphics[trim=0.2cm 0.cm 0.2cm 0.2cm, clip, width=.47\textwidth]{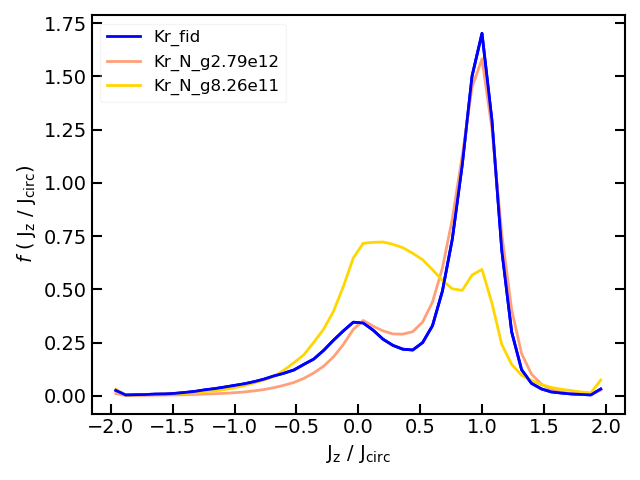} 
\end{minipage} 
\caption{Star formation history {\sl (left-hand panel)} 
and kinematic decomposition of the stellar mass within $0.1$~r$_{\rm vir}$ {\sl (right-hand panel)} 
of galaxy models 
\textcolor{Blue}{\bf Kr\_~fid},  \textcolor{Salmon}{\bf Kr\_N\_g2.79e12}, 
and \textcolor{Goldenrod}{\bf Kr\_N\_g8.26e11}. }
\label{figgAppC} 
\end{figure*}

In this appendix, we generalise our results simulating two different haloes. 
The main goal of this appendix is to investigate how our code performs at simulating haloes expected to host late-type galaxies at redshift $z=0$. For the purpose we adopt other ICs from those of the main paper and therefore we test how our conclusions can be generalized to haloes with different merging history.  
The ICs are drawn from the NIHAO set \citep[e.g.][]{Wang2015, Obreja2016}: we consider 
\textcolor{Salmon}{\bf Kr\_N\_g2.79e12} 
(M$_{\rm halo, \, DM} = 2.79 \times 10^{12}$~M$_{\odot}$ at $z=0$) and 
\textcolor{Goldenrod}{\bf Kr\_N\_g8.26e11} 
(M$_{\rm halo, \, DM} = 8.26 \times 10^{11}$~M$_{\odot}$).
We refer to the NIHAO papers for details about the ICs.  As for gravitational softenings, we adopted the same softening lenghts used for the AqC5 suite (see Section~\ref{sec:MUPPI}). Mass resolution is comparable to that of AqC5 galaxies \citep[see][for details]{Wang2015}.
A comparison between our results and those in the aforementioned papers is beyond the scope of this work. Rather, here we want to show predictions from our fiducial model when different ICs are adopted. 

Figure~\ref{StellarGasMaps_Nihao} introduces stellar and gas projected density maps of galaxy models \textcolor{Salmon}{\bf Kr\_N\_g2.79e12} and 
\textcolor{Goldenrod}{\bf Kr\_N\_g8.26e11} (see the description and discussion of Figure~\ref{StellarGasMaps} for details).

For the same galaxies, we show the SFH and the kinematic decomposition of their stellar mass in the left and right panel of Figure~\ref{figgAppC}, respectively.  
The stellar disc component of \textcolor{Salmon}{\bf Kr\_N\_g2.79e12} is extended and dominant over the bulge 
(B/T$=0.22$; $\text{M}_{\ast} = 1.47 \times 10^{11}$~$\text{M}_{\odot}$). 
On the other hand, the model \textcolor{Goldenrod}{\bf Kr\_N\_g8.26e11} 
has a more prominent stellar bulge 
(B/T$=0.49$; $\text{M}_{\ast} = 2.71 \times 10^{10}$~$\text{M}_{\odot}$). 
Gaseous discs exhibit the clear spiral pattern that the Kr model predicts for the AqC5 galaxies.


\section{Random errors and butterfly effect}
\label{AppD}

\begin{figure*}
\newcommand{\captionfonts}{\small}
\begin{minipage}{\linewidth}
\centering
\includegraphics[trim=0.2cm 0.05cm 0.2cm 0.2cm, clip, width=.52\textwidth]{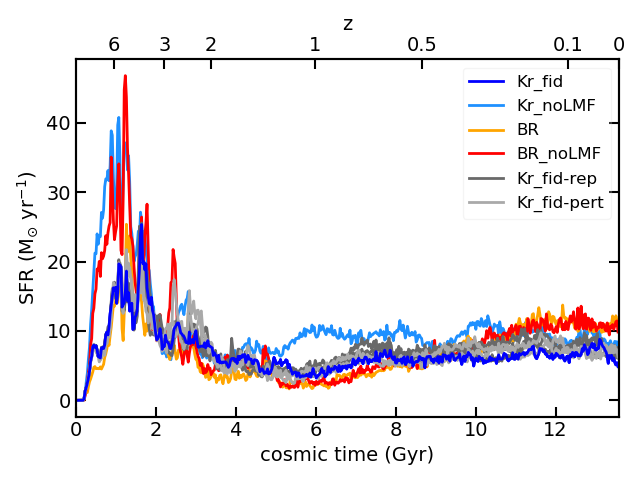} 
\includegraphics[trim=0.2cm 0.cm 0.2cm 0.2cm, clip, width=.47\textwidth]{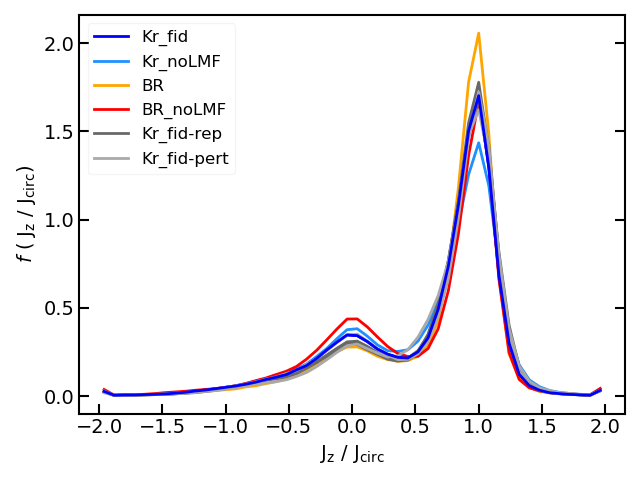} 
\end{minipage} 
\caption{Star formation history {\sl (left-hand panel)} 
and circularity diagram {\sl (right-hand panel)} 
of the galaxy models introduced in Figures~\ref{SFH}~and~\ref{jcirc}. 
In addition, we show repeated runs {\sl (dark grey)} of simulation \textcolor{Blue}{\bf Kr\_~fid},  
and repeated runs wich also feature small perturbations in the ICs {\sl (light grey)}. 
See text for details. }
\label{figgAppD} 
\end{figure*}

In this appendix we address the issue of the possible impact of the chaoticity of the physical system on the final conclusions of our work. Results from cosmological, zoomed-in simulations are affected by the cumulative outcome of a number of uncertainties, e.g.: numerical errors, randomly generated seeds, equations with chaotic solutions 
\citep[see e.g.][]{Genel2019, Keller2019, Oh2020, Davies2021, Granato2021}.  
As a consequence,  predictions from the same simulation repeated several times can differ among them. Despite being often overlooked,  this effect may undermine the robustness of the results.  

To tackle this problem and investigate whether discrepancies produced by the chaotic behaviour of the code are comparable to deviations resulting from different physical assumptions, we proceed as follows.
We consider the fiducial set of our four simulations, namely: \textcolor{Blue}{\bf Kr\_~fid},  \textcolor{Cerulean}{\bf Kr\_noLMF}, 
\textcolor{YellowOrange}{\bf BR}, and \textcolor{Red}{\bf BR\_noLMF}.  
We carry out two pools of simulations of the reference run \textcolor{Blue}{\bf Kr\_~fid}: 
$(i)$ in the first pool, we re-run \textcolor{Blue}{\bf Kr\_~fid} three additional times, always on two different nodes, from identical ICs (for short, repeated runs -- dubbed {\sl rep} in simulation labels); 
$(ii)$ in the second one, we perform two simulations introducing small perturbations in the ICs of \textcolor{Blue}{\bf Kr\_~fid} (perturbed runs -- dubbed {\sl pert} in simulation labels).  
To carry out the second pool of simulations, we randomly displace the initial coordinates of each particle,  independently from one another.  As for the displacement, we draw a number from a uniform distribution spanning the range (-dx, dx), dx being set to $10^{-10} \, \varepsilon_{\rm Pl}$  \citep[as in][]{Granato2021}. 
Perturbations in the ICs of the aforementioned size result in differences in parallel runs that are comparable to adopting different hardware or software set-ups \citep[][]{Keller2019}. 

The left- and righ-hand panels of Figure~\ref{figgAppD} show the predictions from these simulations as for the SFH of simulated galaxies, and the distribution of their stellar mass fraction as a function of the circularity of the orbit. 
Figure~\ref{figgAppD} demonstrates that differences among runs produced by repeated and perturbed runs are not comparable to those driven by different physical processes. We can safely conclude that, within the spread of slightly different evolutions, LMF controls the SFH above $z \gtrsim 1$, while the different prescriptions for the SF are responsible for the SFRs at lower redshift.
A quantification of the impact of the chaotic behaviour of our code is beyond the scope of this work. Here, we qualitatively assess the impact of the numerics, thus proving the robustness of our results.

As for the choice of the reference run \textcolor{Blue}{\bf Kr\_~fid} among the simulations within this set, we simply picked the first run that was initially carried out along with the simulations listed in Table~\ref{tab:sims}, before planning the investigation discussed in this appendix.

\section*{Acknowledgments}
The authors thank the anonymous referee for the careful report that helped improving the presentation of results. 
The authors also thank Pierluigi Monaco and Gabriella De Lucia for interesting discussions and Volker Springel for making an early version of the Gadget3 code available. The authors are grateful to Andrea Macci\`{o} and Aura Obreja for providing the ICs of the NIHAO haloes. 
MV is supported by the Alexander von Humboldt Stiftung and the Carl Friedrich von Siemens Stiftung. MV and KD acknowledge support from the Excellence Cluster ORIGINS, which is funded by the Deutsche Forschungsgemeinschaft (DFG, German Research Foundation) under Germany's Excellence Strategy - EXC-2094 - 390783311. KD acknowledges support by the COMPLEX project from the European Research Council (ERC) under the European Union’s Horizon 2020 research and innovation program grant agreement ERC-2019-AdG 882679.
SB acknowledges financial support from the Istituto Nazionale di Fisica Nucleare (INFN) InDark grant. 
AR acknowledges support from the grant PRIN-MIUR 2017 WSCC32. 
Simulations were carried out using computational resources at the Leibniz Rechenzentrum (LRZ) and at CINECA.

\section*{Data Availability}
The data underlying this article will be shared on reasonable request to the corresponding author.



\bibliographystyle{mnras}
\bibliography{cool_ref}



\bsp	
\label{lastpage}
\end{document}